\documentclass[12pt]{iopart}
\usepackage{graphicx}
\expandafter\let\csname equation*\endcsname\relax
\expandafter\let\csname endequation*\endcsname\relax
\usepackage{amsmath}
\usepackage{amssymb}
\usepackage{makecell}
\usepackage[colorlinks=False]{hyperref}
\begin{document}
\title[Preparation of Rb and Cs in the ground state of an optical tweezer]{Preparation of $^{87}$Rb and $^{133}$Cs in the motional ground state of a single optical tweezer}

\author{S Spence, R V Brooks, D K Ruttley, A Guttridge, and Simon~L~Cornish}

\address{Department of Physics, Durham University, South Road, Durham DH1 3LE, United Kingdom}

\begin{abstract}
\noindent
We report simultaneous Raman sideband cooling of a single $^{87}$Rb atom and a single $^{133}$Cs atom held in separate optical tweezers at 814\,nm and 938\,nm, respectively. Starting from outside the Lamb-Dicke regime, after 45\,ms of cooling we measure probabilities to occupy the three-dimensional motional ground state of 0.86$^{+0.03}_{-0.04}$ for Rb and 0.95$^{+0.03}_{-0.04}$ for Cs.
Our setup overlaps the Raman laser beams used to cool Rb and Cs, reducing hardware requirements by sharing equipment along the same beam path. The cooling protocol is scalable, and we demonstrate cooling of single Rb atoms in an array of four tweezers. After motional ground-state cooling, a 938\,nm tweezer is translated to overlap with a 814\,nm tweezer so that a single Rb and a single Cs atom can be transferred into a common 1064\,nm trap. By minimising the heating during the merging and transfer, we prepare the atoms in the relative motional ground state with an efficiency of 0.81$^{+0.08}_{-0.08}$. This is a crucial step towards the formation of single RbCs molecules
confined in optical tweezer arrays.
\end{abstract}

\noindent{\it Keywords\/}: optical tweezers, merging, ultracold molecules, Raman sideband cooling, array

\section{Introduction}

Optical tweezer arrays trapping individual neutral atoms have been demonstrated as a platform where quantum simulation and quantum information processing could be used to solve complex problems in physics and chemistry \cite{Bergamini2004, Zimmermann2011,Piotrowicz2013,Nogrette2014, Lester2015, Ebadi2021, Scholl2021}.
Dynamic control over trap positions allows deterministic preparation of a fixed number of particles in a scalable geometry \cite{Weiss2004, Barredo2016, Endres2016, Stuart2018}.
Microwave transitions or optical Raman transitions control the internal state of individual particles on timescales much faster than the decoherence time \cite{Young2020}.
Entanglement can be generated within an array by using on-site collisions \cite{Kaufman2015} or dipole-dipole interactions between Rydberg states \cite{Wilk2010, Graham2019, Levine2019}.
Along with high-fidelity readout, the optical tweezer array has all of the essential elements for quantum computation \cite{DiVincenzo2000, Kaufman2021}.
The majority of experiments to date have focused on a single atomic species, although dual-species experiments are emerging \cite{Singh2022, Sheng2022,Zhang2022}. 

Utilising dual-species tweezer arrays opens new avenues for research.
The energy-difference between atomic transitions in different species of atoms allows species-selective loading and imaging \cite{Sheng2022}.
Species-selective imaging could be used to perform quantum non-demolition measurements with low cross-talk \cite{Beterov2015}.
Furthermore, one can create species-selective traps using tweezers with different wavelengths \cite{Singh2022,Brooks2021,Liu2018}.
By merging traps to bring two atoms together, one can produce molecules using microwaves and spin-motion coupling \cite{He2020}, or by using photoassociation \cite{Liu2019}, or by associating across a magnetic Feshbach resonance \cite{Zhang2020}.

Ultracold heteronuclear molecules offer several advantages for experiments in quantum science \cite{Carr2009}, including controllable long-range dipole-dipole interactions, a diverse set of energy levels associated with rotation and vibration, and strong coupling to applied electric and microwave fields. An array of optical tweezers containing single ultracold molecules is an enticing platform for quantum simulation \cite{Micheli2006, Barnett2006, Buchler2007, Gorshkov2013}, quantum computation \cite{DeMille2002, Yelin2006, Zhu2013, Herrera2014, Ni2018, Sawant2020, Hughes2020}, and investigations into ultracold molecular collisions \cite{Krems2008,Christianen2019, Gregory2020, Christianen2021}. New opportunities arise for quantum information processing exploiting the rich internal structure \cite{Sawant2020, Albert2020}. 
All of the aforementioned applications benefit from the preparation of the particle in a single motional state of the tweezer.
For example, reduction in thermal dephasing improves the fidelity of the transfer between quantum states \cite{Kuhr2005, Kaufman2012, Lorenz2021}, an essential component for implementing quantum gates.
Single atoms can be prepared in the ground state of the harmonic trapping potential through the process of Raman sideband cooling (RSC) \cite{Kaufman2012, Thompson2013, Sompet2017, Yu2018, Wang2019,Lorenz2021}. 
Certain species of molecule can be directly cooled \cite{Barry2014, Zhelyazkova2014, Anderegg2018, Collopy2018}
and have recently been trapped in optical tweezers \cite{Anderegg2019}. However, cooling to the motional ground state is an ongoing challenge due to the complex energy level structure of molecules \cite{Caldwell2020}.
Alternatively, if a pair of atoms in the relative motional ground state are magnetoassociated into a molecular state  \cite{Thalhammer2006, Ospelkaus2006, Zhang2020}, the molecule will inherit the initial motional state {of the atom pair} \cite{He2020, Zhang2020}.
In fact, the initial preparation of the relative motional ground state is essential for the efficient magnetoassociation of the two atoms into a molecule \cite{Busch1998, Zhang2020}.  
This approach has recently been used to prepare a tweezer array of NaCs molecules in their motional ground states \cite{Zhang2022}.
Extending this approach to other bialkali molecules will open up new opportunities, leveraging the existing work on bulk gases. In the case of RbCs molecules, the rotational and hyperfine structure \cite{Aldegunde2008, Gregory2016, Blackmore2020b}, and AC stark shifts \cite{Gregory2016, Blackmore2020} have already been characterised in great detail. This has enabled the experimental demonstration of robust storage qubits based upon long-lived coherent superpositions of hyperfine states \cite{Gregory2021}. Moreover, the molecular structure of the $X^1\Sigma^+ \rightarrow b^3\Pi$ transition in RbCs has been shown to be ideally suited for the construction of a magic trap for multiple rotational transitions \cite{Guan2021}, permitting long rotational-state coherences and opening up interesting possibilities to encode synthetic dimensions in the molecule \cite{Sundar2018}.

This work outlines progress towards creating an array of ultracold $^{87}$Rb$^{133}$Cs molecules.
We demonstrate simultaneous RSC of a single rubidium ($^{87}$Rb) atom and a single caesium ($^{133}$Cs) atom held in separate optical tweezers.
We achieve a probability of {0.86$^{+0.03}_{-0.04}$ }
that a Rb atom occupies the motional ground state of an 814\,nm optical tweezer.
Similarly, we achieve a probability of 0.95$^{+0.03}_{-0.04}$
that a Cs atom occupies the motional ground state of a 938\,nm tweezer. 
{Our cooling protocol is designed to cool atoms from outside the Lamb-Dicke regime in order to access the benefits of using moderate tweezer powers. The challenge of cooling in a relatively shallow trap is overcome using higher order sideband transitions \cite{Yu2018}.}
Furthermore, we show that the cooling is possible using a condensed optical setup that shares hardware between the RSC laser beam paths for both species.
We demonstrate the scalability {of our setup} by simultaneously cooling {four Rb atoms in a one-dimensional tweezer array} to a mean motional ground-state probability of $0.64^{+0.03}_{-0.05}$.
Finally, following cooling, a Rb and a Cs atom are transferred with minimal heating to the same trapping potential of a 1064\,nm tweezer. 
We achieve a probability of 0.81$^{+0.08}_{-0.08}$ for preparing the atoms in their relative motional ground state in the Zeeman hyperfine states $|f_{\mathrm{Rb}}=2,m_{f,\mathrm{Rb}}=2\rangle$ and $|f_{\mathrm{Cs}}=4, m_{f,\mathrm{Cs}}=4\rangle$. 

The structure of this paper is as follows.
To begin, section~\ref{sec:rsc_method} describes the general protocol for RSC.
Section~\ref{sec:experimental_setup} provides a description of our experimental setup.
Section~\ref{sec:pulse_sequence} outlines the theoretical model we use to design the optimal  cooling sequence, before presenting the design of a pulse sequence that reaches the motional ground state having started outside the Lamb-Dicke regime.
In section~\ref{sec:separate_tweezers}, sideband thermometry is used to quantify the ground-state occupation after RSC and demonstrate high-fidelity preparation of atoms in the motional ground state, including results from cooling an array.
Finally, section~\ref{sec:merging} reports the optimisation and performance of the merging sequence used to prepare a Rb-Cs atom pair in the motional ground state of a common 1064\,nm tweezer.

\section{Method of Raman Sideband Cooling}
\label{sec:rsc_method}

\begin{figure}
    \centering
    \includegraphics[width=1\linewidth]{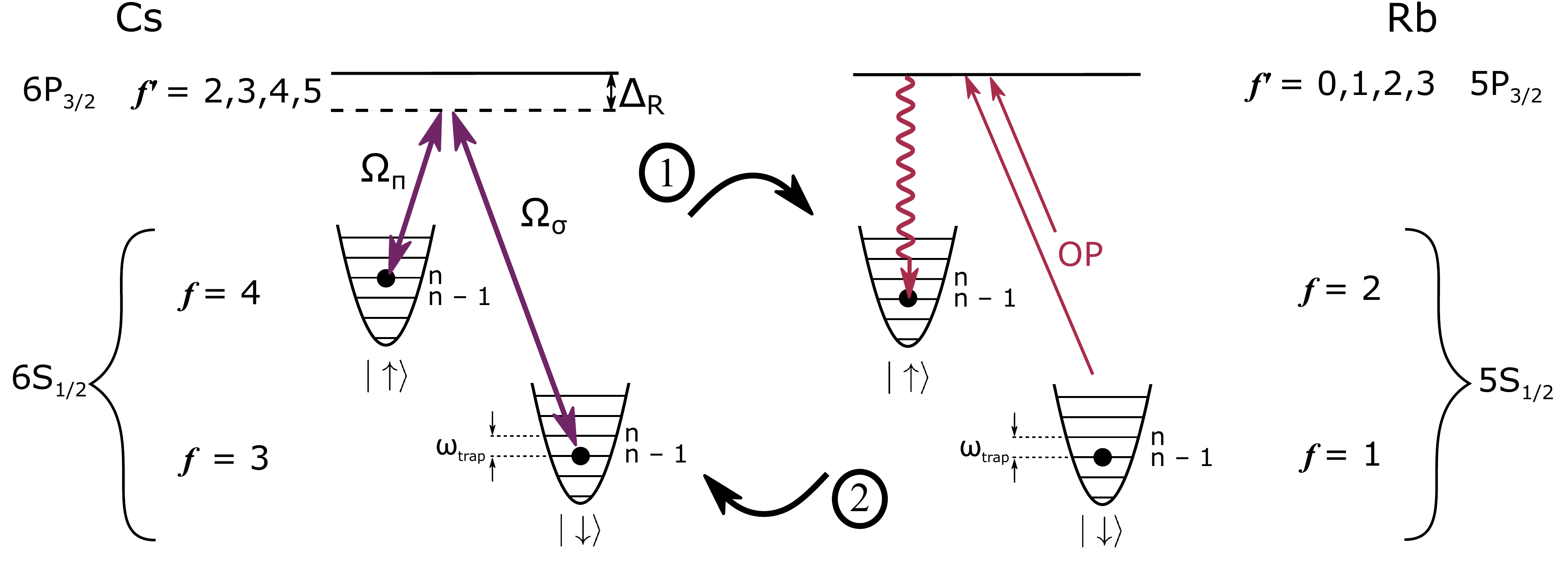}
    \caption{The two stages of a Raman sideband cooling iteration. First a coherent two-photon Raman transition transfers some of the population from $|\mathord{\uparrow}; n\rangle \rightarrow |\mathord{\downarrow}; n-1\rangle$. One beam is circularly polarised for $\sigma^+$ transitions with Rabi frequency $\Omega_{\sigma}$. The other is linearly polarised for $\pi$ transitions with Rabi frequency $\Omega_{\pi}$. The single-photon detuning from the excited state is $\Delta_{\mathrm{R}}$. Then a dissipative optical pumping step resets the spin, preserving the motional level: $|\mathord{\downarrow}; n-1\rangle \rightarrow |\mathord{\uparrow}; n-1\rangle$. Each iteration of these stages removes one quanta of motional energy, $\hbar \omega_\mathrm{trap}$, where $\omega_\mathrm{trap}$ is the trap frequency. 
    The hyperfine spin state manifolds are labelled by the total angular momentum quantum number, $f$. 
    }
    \label{raman_transitions}
\end{figure}

RSC relies on two processes to transfer the atom between the motional Fock states $|n\rangle$ of the optical tweezer trap.
In the first step, a stimulated two-photon Raman transition transfers the atom between hyperfine spin states, $|\mathord{\uparrow}\rangle$ and $|\mathord{\downarrow}\rangle$. 
When the transition is on resonance with a lowering sideband, it performs a spin flip and reduces the motional level: $|\mathord{\uparrow}; n\rangle \rightarrow |\mathord{\downarrow}; n-1\rangle$.
Then, in the second step, optical pumping (OP) transfers the population back into the original hyperfine state while preserving the motional level: $|\mathord{\downarrow}; n-1\rangle \rightarrow |\mathord{\uparrow}; n-1\rangle$.
The combination of these two processes reduces the motional level by one quanta, as illustrated in Fig.~\ref{raman_transitions}.
Iterating over the procedure cools the atom into the lowest motional level, at which point there is no further level to descend to, and so the atom decouples from both the Raman and the OP light.

The tight confinement of optical tweezers puts the atom in the Lamb-Dicke (LD) regime, allowing control over the motional level through atom-light interactions.
Atoms are illuminated by laser light, leading to photon scattering events which result in atomic recoil due to the conservation of momentum.
The LD parameter $\eta  =\sqrt{\omega_{\mathrm{recoil}}/\omega_{\mathrm{trap}}} = \sqrt{\hbar k^2/(2m \omega_{\mathrm{trap}})}$ is determined by the trap frequency, $\omega_\mathrm{trap}$, and the photon recoil energy, $\hbar\omega_\mathrm{recoil}=\hbar^2k^2/2m$ for resultant wavevector $k$ and mass $m$. The LD parameter satisfies $\eta^2 (n+1) \ll 1$ in the LD regime, resulting in a suppression of motional excitation during photon scattering events \cite{Wineland1998}.
Being in the LD regime is important for both of the aforementioned steps of RSC.
In the OP step, the excitations and subsequent spontaneous emissions are on the carrier transition, i.e. they preserve the motional level.
But it is also possible to make transitions between specific motional states - sideband transitions - provided the transition linewidth is smaller than the spacing of the energy levels.
A stimulated two-photon Raman transition satisfies this condition by coupling two long-lived states via an excited state that is not populated \cite{Wu1996, Bateman2010}.
Sideband transitions that change the motional level occur at intervals of the trap frequency.
In standard notation, a blue sideband (BSB) transition increases the motional level, whereas a red sideband (RSB) transition reduces the motional level. 
The direction of the atomic recoil momentum determines which trap axes the Raman transition can couple to.
To cool to the 3D ground state, the laser beams driving Raman transitions must be arranged to allow coupling to the different trap axes.

The effectiveness of an RSC protocol is determined by the competition between cooling rates and heating rates.
The main limitation on the cooling rate is the reduced sideband transfer due to dephasing from differential light shifts, beam power fluctuations, and magnetic field noise. 
The important sources of heating are intensity and pointing noise from the tweezer trap, recoil from OP photons, and off-resonant {carrier and} BSB transitions.
These obstacles to effective ground-state cooling are addressed in section~\ref{sec:pulse_sequence}.

\section{Experimental Setup}
\label{sec:experimental_setup}

\begin{figure}
    \centering
    \includegraphics[width=1\linewidth]{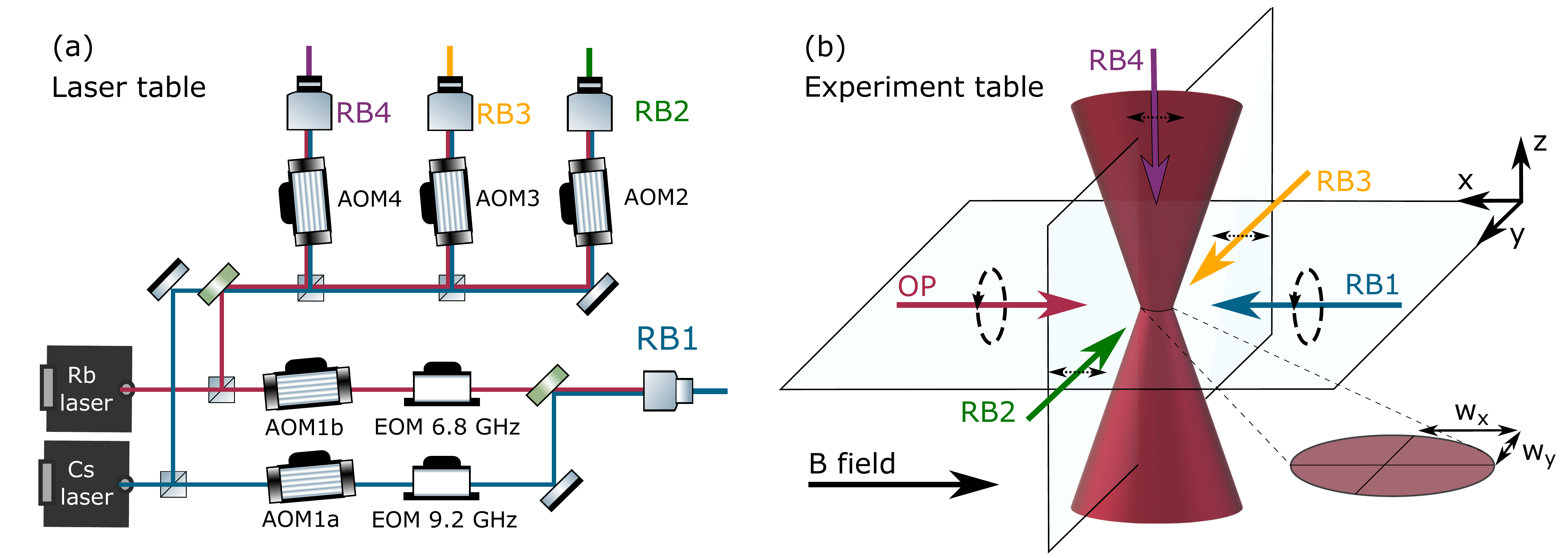}
    \caption{(a) Lasers operating at 780\,nm and 852\,nm generate light for driving Raman transitions in Rb and Cs, respectively. Electro-optic modulators (EOMs) add frequency sidebands to the RB1 beams at the hyperfine splitting of the electronic ground-state, prior to overlap. For the remaining Raman beams, the 780\,nm and 852\,nm light is first overlapped and then split into three separate beam paths, each with an acousto-optic modulator (AOM) to control the power. (b) Raman beams are fibre-coupled to the main experiment and focused down to beam waists of $100-160\,\mu$m at the position of the atoms. RB1 drives $\sigma^+$ transitions and is pulsed on with one of the other linearly polarised Raman beams to give a two-photon transition coupling the harmonic motion along one of the trap axes. We use RB1+RB4 to couple to the atomic motion along the propagation direction of the tweezer, z. Both RB1+RB2 and RB1+RB3 can couple to the motion in both radial directions. We use RB1+RB2 for the x-direction, and RB1+RB3 for the y-direction. The inset depicts the radial asymmetry where the beam waists satisfy $w_\mathrm{x}>w_\mathrm{y}$. The optical pumping beam is aligned with the quantisation axis to drive $\sigma^+$ transitions with high fidelity. 
    }
    \label{geometry}
\end{figure}

Here we give an overview of the experimental setup for RSC by outlining a typical cooling sequence.
The most relevant components are displayed in Fig.~\ref{geometry}.
The initial preparation of Rb and Cs in species-selective tweezers is described in detail in Ref.~\cite{Brooks2021}.

A typical cooling sequence begins with loading a Rb atom from a magneto-optical trap (MOT) into an optical tweezer with a wavelength of 814\,nm, and subsequently loading a Cs atom from a MOT into a tweezer with wavelength 938\,nm.
This is a stochastic process that takes $\sim$250 ms, after which each tweezer has $\sim$50\% probability of being occupied by a single atom.
The tweezers are formed by focusing the laser beams through the same high numerical aperture objective lens.
Their foci are not radially symmetric due to clipping of the laser beams before they enter the objective lens.
At the focus the 938\,nm tweezer has a beam waist $\{w_{\mathrm{x}}^\mathrm{938}, w_{\mathrm{y}}^\mathrm{938}\} = \{1.29(4), 1.06(2) \}~\mu$m, and the 814\,nm tweezer has a beam waist $\{w_{\mathrm{x}}^\mathrm{814}, w_{\mathrm{y}}^\mathrm{814}\} = \{1.03(3), 0.83(2) \}~\mu$m.
When the atoms are loaded into the tweezers, they are separated by a distance of 4.5\,$\mu$m in the x-direction. We use fluorescence imaging to detect the occupation of the tweezers.
Initially, we use a release and recapture technique \cite{Tuchendler2008} to measure the temperature of the atoms in the tweezers. 
\label{change1}This temperature corresponds to the mean energy of the atom's radial motion after averaging over many iterations of the experiment, where each iteration samples motional levels in the x- and y-direction from independent thermal distributions. 
After 10 ms of polarisation gradient cooling, we typically measure a temperature of $\sim$15\,$\mu$K for Cs in a 1.3~mK deep 938\,nm trap, and $\sim30\,\mu$K for Rb in a 1.3~mK deep 814\,nm trap.
Following the initial cooling, the powers of both tweezers are increased so that the trap depths are 2~mK for Cs and 1.5~mK for Rb. 
The typical trap frequencies are \{$\nu_{\mathrm{x}}, \nu_{\mathrm{y}}, \nu_{\mathrm{z}}\}^{938}_\mathrm{Cs} = \{84, 120, 17\}$\,kHz and \{$\nu_{\mathrm{x}}, \nu_{\mathrm{y}}, \nu_{\mathrm{z}}\}^{814}_\mathrm{Rb} = \{107, 163, 25\}$\,kHz.
At this point typical mean motional levels are $\{n_{\mathrm{x}}, n_{\mathrm{y}},n_{\mathrm{z}}\} \approx \{2, 1, 10\}$.
Then RSC pulses of 780\,nm light for Rb and 852\,nm light for Cs are applied simultaneously to cool the atoms to the motional ground state.
In order to detect the mean motional state, a Raman pulse is used to transfer population out of the upper hyperfine level. 
Finally, a resonant push-out pulse ejects any atom remaining in the upper hyperfine level and hence maps the atomic spin state onto the trap occupancy in a second fluorescence image \cite{Kuhr2003, Jones2007, Brooks2021}. 

To perform efficient RSC it is crucial to maintain the motional state while changing the spin state. A limiting factor is the spin-motion coupling introduced by the vector light shift of the optical tweezer trap \cite{Dareau2018, He2020}. For a linearly polarised tweezer, the tight focusing of the light introduces some ellipticity to its polarisation around the focus, resulting in a vector light shift equivalent to a nonuniform fictitious magnetic field \cite{Thompson2013, Kaufman2012, Albrecht2016}. 
As the vector light shift has a spatial dependence, there is an effective magnetic field gradient that offsets the trap centre for different $m_f$ states. 
The displacement in the trap centre for a spin flip with $\Delta m_f = 1$ between hyperfine states is similar to the ground-state atomic wavepacket size. Therefore, there is a high probability of motional excitation during the OP step.
In our setup, the tweezer polarisation is set along the x-axis which results in an effective magnetic field that points in the y-direction. We suppress the effective field gradient, and consequently the spin-motion coupling, by applying a magnetic field of $4.8$~G along the x-direction during OP and RSC which is perpendicular to the effective magnetic field \cite{Thompson2013}. This applied magnetic field contributes $\lesssim1$~mG of magnetic field noise and we measure drifts in the ambient field of order $1$~mG from day-to-day. We will compare the dephasing on Raman transitions from vector light shifts and magnetic field noise in section~\ref{sec:separate_tweezers}.

It is also important to have high-fidelity state preparation and OP during RSC. 
The same laser beams are used for both tasks.
Our scheme pumps to a spin-stretched hyperfine sub-level with total angular momentum quantum number $f$ using resonant excitation on the D$_2$ line with circularly polarised light driving $\sigma^+$ transitions.
To ensure that there is only a single dark state, we use two overlapped laser beams for each atomic species, derived from the lasers used for the MOT beams.
The first beam drives $|f=i-1/2, m_f\rangle \rightarrow |f'=i+1/2, m_{f'}=m_f+1\rangle$ transitions, where the nuclear spin quantum number is $i=3/2$ for Rb and $i=7/2$ for Cs.
The second beam drives $|f=i+1/2, m_f\rangle \rightarrow |f'=i+1/2, m_{f'}=m_f+1\rangle$ transitions.
In this case, the spin-stretched states, $|f=2,m_{f}=2\rangle$ for Rb and $|f=4,m_{f}=4\rangle$ for Cs, are dark to the OP light provided the polarisation is pure.
High-purity circular polarisation is achieved using a polariser with extinction $>$~5000~:~1 followed by an achromatic quarter waveplate.
We measure a polarisation purity of $>2000:1$ using the atoms after optimising the angle of the quarter waveplate and the direction of the bias field.

Fig.~\ref{geometry} depicts the lasers used for Raman transitions.
All of the Raman beams for one species are generated from the same laser, which ensures the phase coherence of two-photon Raman transitions. 
Our choice of laser frequency is a compromise between reducing off-resonant single-photon scattering whilst maintaining sufficiently strong coupling for two-photon Raman transitions. 
The 780\,nm laser is red-detuned with a single-photon detuning of $\Delta_{\mathrm{R}}=50$~GHz from the Rb D$_2$ line, and the 852\,nm laser is red-detuned $\Delta_{\mathrm{R}}=41$~GHz from the Cs D$_2$ line. 

Fig.~\ref{geometry}(a) displays the hardware controlling the Raman beams. 
The 852\,nm light is overlapped with the 780\,nm light using a dichroic mirror so that they share hardware in the RB2, RB3, and RB4 beam paths.
Frequency and power control is achieved using five acousto-optic modulators (AOMs).
The Bragg diffraction angle of each AOM is wavelength dependent, so we must compromise the diffraction efficiency between the optimum for 780\,nm and the optimum for 852\,nm.
Despite this, we typically achieve first order diffraction efficiencies of $>$~50\% for both wavelengths.
Electro-optic modulators (EOMs) are used to add frequency sidebands at 6.8~GHz for Rb and 9.2~GHz for Cs.
To suppress the possibility of driving unwanted transitions and introducing pathways for quantum interference \cite{Kerman2002}, we offset the EOM frequency by 10\,MHz from the Zeeman-shifted ground-state hyperfine splitting \cite{Kaufman2012} (see \ref{appendix:dme}). 
The use of separate AOMs for RB1 (AOM1a for Cs and AOM1b for Rb) allows for independent control over the two-photon detuning and Rabi frequency of each species.
However, sharing AOMs means that the pulse durations of RB2, RB3, and RB4 are constrained to be the same for both wavelengths.
We therefore set the Rabi frequencies such that the sideband $\pi$-pulse duration is equal for Rb and Cs. 

The geometry of Raman beams in Fig.~\ref{geometry}(b) allows us to couple to the motional state along the three orthogonal axes of the trap.
The two-photon Raman transitions are also coupling the spin states $\{|\mathord{\downarrow}\rangle, |\mathord{\uparrow}\rangle\}= \{|f=3,m_{f}=3\rangle, |4,4\rangle\}$ for Cs, or $\{|\mathord{\downarrow}\rangle, |\mathord{\uparrow}\rangle\}=\{|1,1\rangle, |2,2\rangle\}$ for Rb.
To couple to the motion along a given axis, the resultant wavevector from the combination of the two beams must have a non-zero projection along that trap axis.
The tweezer propagates along the z-direction and the radial asymmetry in the x-y plane is depicted in the inset of Fig.~\ref{geometry}(b).
It should be noted that both RB1+RB2 and RB1+RB3 are capable of coupling to both radial axes.
The trap asymmetry means sideband transitions along both radial axes can be spectrally resolved.
Strictly speaking, only one of RB2 or RB3 is required. Although the current work makes use of both RB2 and RB3, we have verified that the cooling protocol reaches the same final ground-state fraction using only RB2.
The Raman beams are focused onto the atoms to produce beam waists of $100-160\,\mu$m so that modest laser powers can be used to achieve the desired Rabi frequencies. {We note that a combined Raman beam power of 2~mW at the atoms is sufficient for our cooling scheme.}
The Rabi frequency is chosen to maximise the cooling rate, but is constrained by other parameters such as the trap frequency, as will be explained in the next section.

\section{Designing the Protocol for Raman Sideband Cooling}
\label{sec:pulse_sequence}
Our RSC protocol uses a sequence of Raman pulses to cool an atom to the motional ground state, starting from outside the LD regime. In order to accomplish this, the Raman pulses are shaped with a smoothed temporal profile and the pulse sequence targets several different sideband transitions. 
Using these techniques enables cooling at lower trap depths where both the decoherence caused by differential light shifts and the heating from photon scattering are reduced.
More specifically, lowering the trap depth can increase the effectiveness of RSC by increasing the cooling rate and reducing heating rates.
The cooling rate is proportional to the sideband transfer efficiency, which is improved by reducing dephasing from differential light shifts that scale proportional to the trap depth \cite{Kuhr2005, Thompson2013}.
The differential light shifts are significant in our case because the tweezers are near-detuned to ensure species-selectivity \cite{Brooks2021}.
Furthermore, the heating rates from tweezer photon scattering, intensity noise, and pointing noise are reduced by using lower trap depths \cite{Savard1997, Gehm1998}.
Finally, given the practical limit of maximum tweezer power available, using less power per trap allows the production of larger arrays.
Therefore there is high incentive to design a cooling protocol that works for shallow traps where the atom is initially outside the LD regime. {Below we provide more details of the measures we have implemented to overcome the challenges of implementing RSC in relatively shallow traps.}

\begin{figure}
    \centering
    \includegraphics[width=1\linewidth]{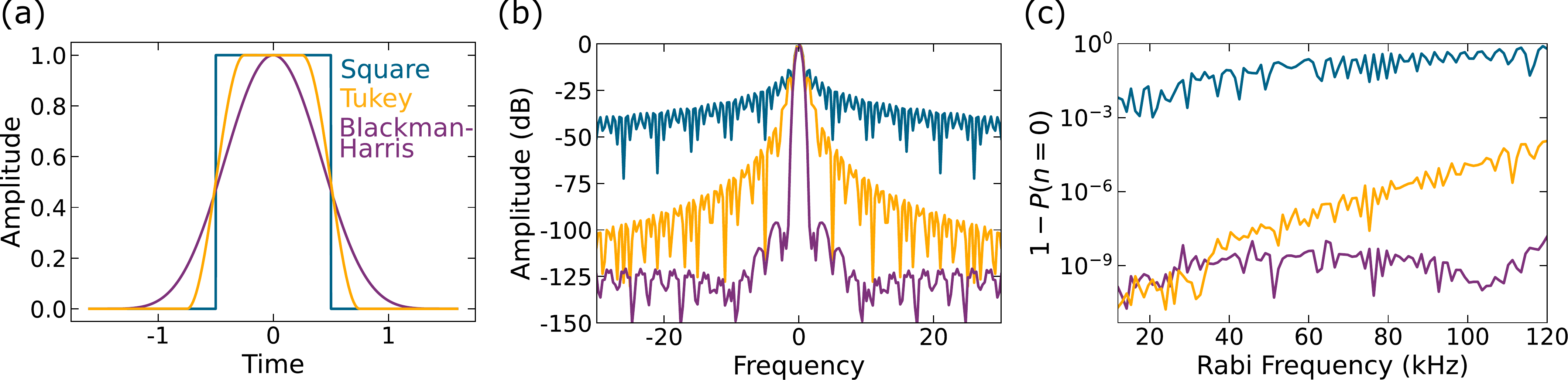}
    \caption{
    {The temporal and spectral profiles of different pulse shapes.}
    (a) The Tukey (with a cosine fraction of $2/3$) and Blackman-Harris pulse shapes have the same pulse area as a square pulse if the durations are 1.5 and 2.79 times longer respectively and the peak amplitude is the same.
    (b) The Fourier spectrum of a square pulse demonstrates broad sidelobes. The temporal smoothing of the Tukey and Blackman-Harris pulse shapes suppresses the sidelobes.
    (c) Starting in $|n=0\rangle$, the probability of excitation after applying a Raman pulse on the RSB increases with the Rabi frequency, $\Omega_\mathrm{R}$. The peak Rabi frequency is scaled to keep the pulse area the same for the different pulse profiles.
    The pulse shapes with a broad Fourier spectrum have a higher probability of excitation. The trap frequency is $\omega_\mathrm{trap}=120$\,kHz and the pulse duration is $T=\pi/\eta\Omega_\mathrm{R}$, with LD parameter $\eta=0.13$. 
    }
    \label{pulse_shapes}
\end{figure}

\subsection{Pulse shaping}
Pulse shaping is required when using lower trap depths to {reduce} the probability of off-resonant Raman carrier or BSB transitions.  Lowering the trap depth decreases the sideband splitting such that the finite width of the Raman pulse's Fourier spectrum results in an increased probability of off-resonant excitation.
These undesired transitions reduce the cooling rate by skipping a cooling step and necessitating another OP step to reset the spin, which has associated heating from photon recoil.

A suitable pulse shape is chosen based on the balance between its {spectral width and the temporal duration required to achieve a $\pi$-pulse.} 
Fig.~\ref{pulse_shapes}(a) compares a square pulse profile to a Blackman-Harris profile and a Tukey profile with cosine fraction 2/3 \cite{Harris1978}.
The first consideration is to minimise the spectral width to avoid off-resonant excitation from two-photon Raman transitions.
The abrupt change in amplitude of the square pulse results in the appearance of sidelobes in the Fourier spectrum in Fig.~\ref{pulse_shapes}(b).
The Tukey profile smooths the edges of the pulse to suppress sidelobes.
The Blackman-Harris profile is shaped to minimise the sidelobes in the Fourier spectrum.
Fig.~\ref{pulse_shapes}(c) displays the results of solving the Schr\"{o}dinger equation for the application of a Raman $\pi$-pulse on the RSB, having started in the motional ground state.
Here we consider the radial direction, using a trap frequency of $\omega_\mathrm{trap}=120$\,kHz and LD parameter $\eta=0.13$.
The square pulse profile has a significant probability of off-resonant excitation. 
The excitation probability is reduced by using a Tukey profile, but still increases with the Rabi frequency.
For our radial trap frequencies, the Tukey profile maintains the minimal excitation given by the Blackman-Harris profile provided that the Rabi frequency is $< 40$\,kHz.
The side effect of smoothing the pulse profile is a lower mean Rabi frequency; in order to achieve the same pulse area either a longer duration or a higher peak Rabi frequency is required.
Longer pulses are undesirable as they allow time for spontaneous scattering from the tweezer or the Raman beams. 
And given that there is limited laser power available for the Raman beams, we choose the Tukey profile for the radial directions so that a lower peak Rabi frequency is required.
However, in the axial direction the smaller trap frequency necessitates using a Blackman-Harris profile.
In practice, RB1 is always a square pulse, so the Raman coupling is the convolution of the RB2, RB3 or RB4 pulse profile with a square pulse.
In the radial directions, the resultant square-root Tukey profile still maintains an acceptable excitation probability of $< 10^{-5}$ for Rabi frequencies $< 30$\,kHz.
However, for the axial direction we shape the RB4 pulse with the square of a Blackman-Harris profile, such that the convolution of RB1+RB4 is a Blackman-Harris profile.

\begin{figure}
    \centering
    \includegraphics[width=1\linewidth]{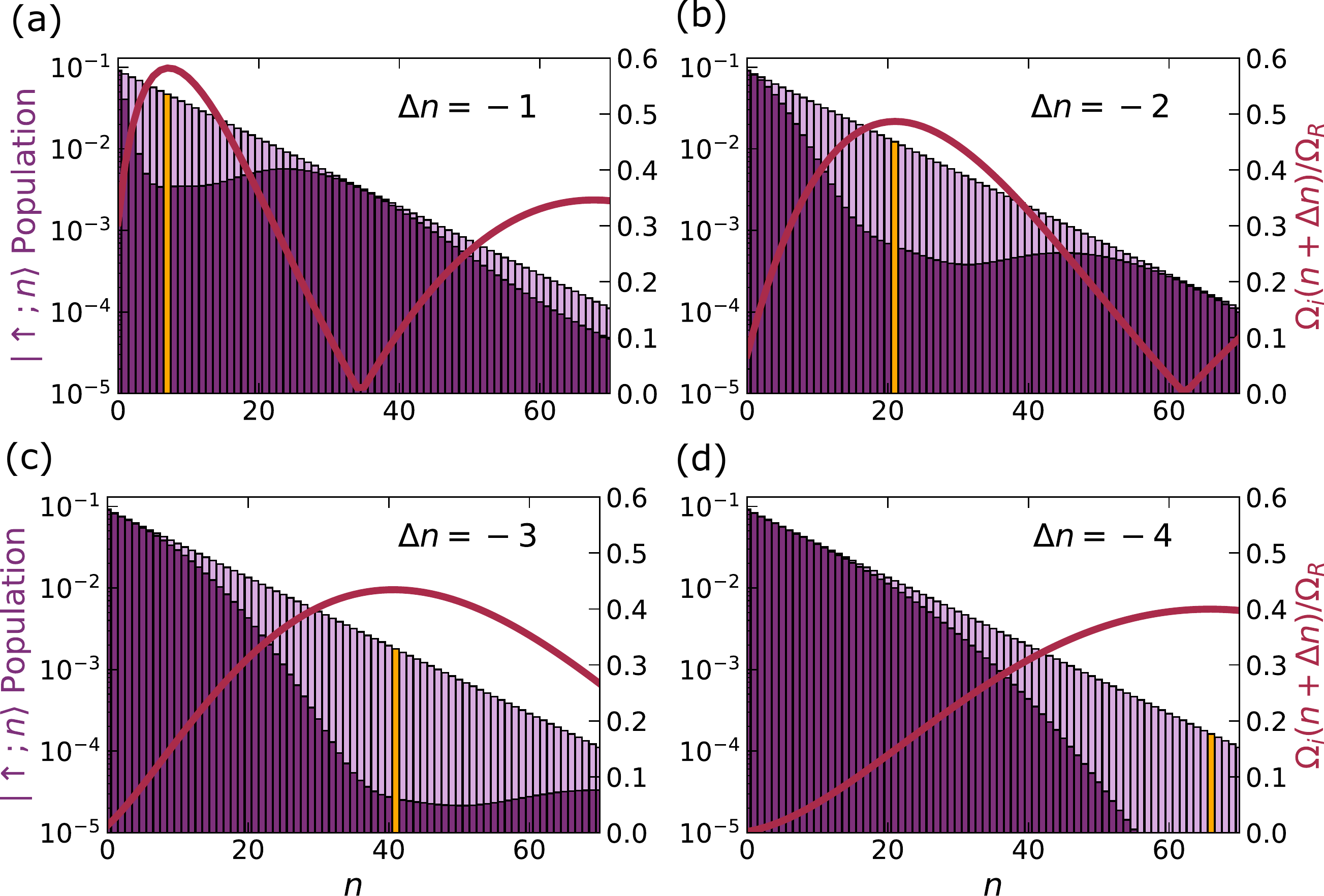}
    \caption{Outside of the LD regime, higher-order sideband transitions enable coupling to the full thermal distribution of motional levels.
    A simulation of applying an $n \rightarrow n+\Delta n$ sideband $\pi$-pulse for the motional level, {$n_{\mathrm{max}}$}, that maximises the sideband Rabi frequency: $T = \pi/\Omega_i(n_{\mathrm{max}},n_{\mathrm{max}}+\Delta n)$ where $d\Omega_i(n_{\mathrm{max}},n_{\mathrm{max}}+\Delta n)/dn = 0$, with a LD parameter $\eta=0.35$. Lilac: the initial thermal population distribution for the $|\mathord{\uparrow}\rangle$ spin state with $\langle n\rangle =10$. Purple: The population distribution after applying the Raman pulse. 
    {Gold: $n_\mathrm{max}$}. Red curve: the sideband Rabi frequency $\Omega_i$ normalised by the peak Raman Rabi frequency $\Omega_{\mathrm{R}}$ demonstrates the variation of coupling as a function of the initial and final motional levels: $\Omega_i(n+ \Delta n) / \Omega_{\mathrm{R}}= |\langle n|e^{j\eta_i (a+a^\dagger)}|n+\Delta n\rangle |$. (a) $\Delta n=-1$, $n_{\mathrm{max}} = 7$. (b) $\Delta n=-2$, $n_{\mathrm{max}} = 21$. (c) $\Delta n=-3$, $n_{\mathrm{max}} = 41$. (d) $\Delta n=-4$, $n_{\mathrm{max}} = 66$.}
    \label{higher_sideband_pulses}
\end{figure}

\subsection{Sideband Transitions Outside the Lamb-Dicke Regime}

Using shallow trap depths means that the atom starts outside the LD regime, such that the Raman coupling depends on the motional level.
The rate at which population is transferred between motional states is defined by the Rabi frequency \cite{Schwabl1995}:
\begin{equation}
	\Omega_{i}(n,m) = \Omega_{\mathrm{R}} e^{-\eta_{i}^2/2} \sqrt{\frac{n_{<}!}{n_{>}!}}\eta_{i}^{|n-m|} L^{|n-m|}_{n_{<}}(\eta_{i}^2).
    \label{eq:1}
\end{equation}
Here $n_{<}$ is the smaller of the motional levels $\{n, m\}$, and $n_{>}$ is the larger.
The Raman Rabi frequency, $\Omega_{\mathrm{R}} = \Omega_\pi \Omega_\sigma/(2\Delta_{\mathrm{R}})$, depends on the single-photon Rabi frequencies, $\Omega_\pi$ and $\Omega_\sigma$, and the single-photon detuning from the P$_{3/2}$ manifold, $\Delta_{\mathrm{R}}$.
$L^{|n-m|}_{n_{<}}(\eta^2)$ is an associated Laguerre polynomial \cite{Arfken1985}.
The LD parameter, $\eta_i$ for trap axis $i\in$\{x, y, z\}, is dependent on the beam geometry that determines the direction of atomic recoil.
In our setup the recoil momentum is always at 45 degrees to the trap axis that it couples to.
In the LD regime, the Rabi frequency has only a weak dependence on the motional level.
However, in the axial direction where $\eta_\mathrm{z}\sim$ 0.3, the initial thermal distribution with $\langle n \rangle \sim 10$ starts outside of the LD regime. 
Fig.~\ref{higher_sideband_pulses}(a) shows the result of solving the Schr\"{o}dinger equation for the application of a Raman pulse on the first RSB, demonstrating population trapping where the Raman coupling vanishes at $n=35$.
In contrast, Fig.~\ref{higher_sideband_pulses}(b)-(d) demonstrate how higher order sideband transitions can be used to achieve strong coupling to motional levels of $n>20$.
In order to address a wide range of motional states, the pulse duration is set to a $\pi$-pulse for $n_\mathrm{max}$, the motional level that maximises the sideband Rabi frequency: $T = \pi/\Omega_i(n_{\mathrm{max}},n_{\mathrm{max}}+\Delta n)$ where $d\Omega_i(n_{\mathrm{max}},n_{\mathrm{max}}+\Delta n)/dn = 0$. 
Cooling from outside the LD regime is achieved using a sequence of pulses targeting different sideband transitions.

\begin{figure}
    \centering
    \includegraphics[width=1\linewidth]{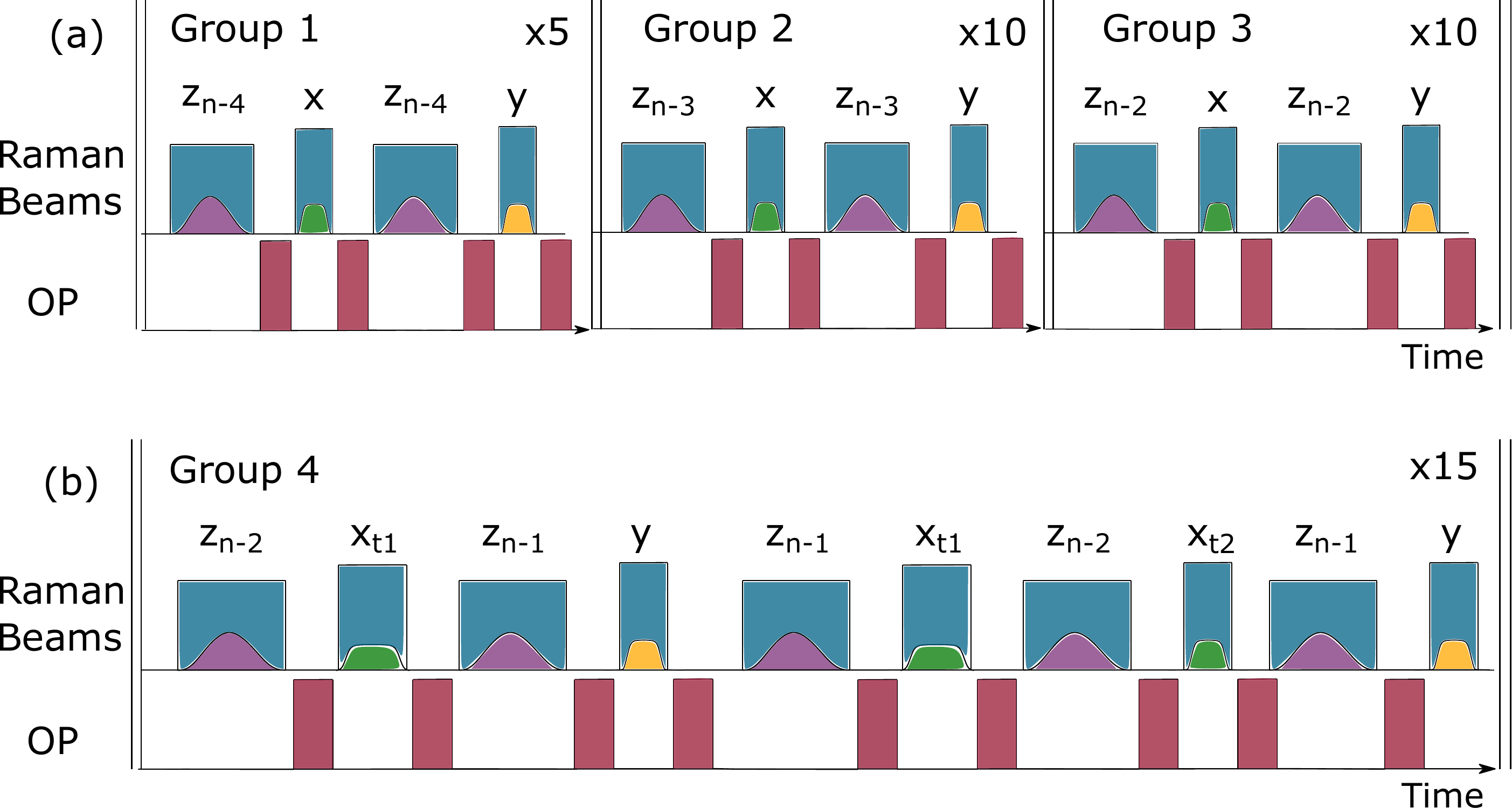}
    \caption{
    {{Overview of the final RSC pulse sequence.} The sequence is split into groups that are repeated, and each group alternates pulses between a pair of Raman beams and an OP step. 
    The labels denote the axis {(x,y,z) addressed by each beam pair. The subscripts on z denote which axial sideband transition is driven}. (a) Groups 1, 2, and 3 have the same format. }
    RB1 (blue) uses a square pulse, whereas RB2 (green) and RB3 (yellow) use a Tukey pulse shape and RB4 (purple) uses the square of a Blackman-Harris pulse shape. 
    (b) {The pulse structure used for group 4 of the full sequence. Here we target the $n-2$ and $n-1$ sidebands in the axial direction}, and the inner radial direction (x) uses two different durations denoted by the subscripts t1 and t2. }
    \label{pulse_sequence}
\end{figure}

\subsection{Assembling a Pulse Sequence}

We {use simulations to} model different pulse sequences and guide the choice of {several} experimental parameters relevant to the cooling scheme. 
We do so by solving the Lindblad master equation for the evolution of the atomic state during a RSC pulse sequence \cite{Cirac1994, Kaufman2012}.
The simulations include the coherent transfer from the Raman beams, and dissipative OP steps.
The methodology is outlined in \ref{appendix:simulation}.
Even without considering other heating effects such as scattering from the tweezer, we are able to draw several conclusions.
In the radial direction, using a fixed {pulse} duration achieves a similar fidelity of motional ground-state preparation {compared} to an optimised routine where each pulse duration is an independent variable \cite{Rasmusson2021}.
However, in the axial direction, achieving high fidelity motional ground-state preparation requires addressing the motional levels $n>20$ using sideband transitions with $|\Delta n| > 1$.
Our simulations showed that the higher motional levels should be addressed first, bunching the population distribution in the lower motional levels.

Randomising the two-photon detuning between time steps allows us to simulate drifts in two-photon detuning and places limits on the optimal choice of Rabi frequency.
While reducing the Rabi frequency reduces the probability of off-resonant Raman transitions, it also reduces the width of the sideband transition and reduces sideband transfer in the presence of dephasing.
The consequence is increased sensitivity to changes in the two-photon detuning. 
The increased $\pi$-pulse duration is also detrimental due to the photon scattering and trap heating effects discussed earlier.
Therefore, guided by simulations, we compromise by choosing mean Rabi frequencies of 4\,kHz for the axial direction and 20-30\,kHz for the radial directions.

Finally, we bring the previous considerations together to construct a pulse sequence {that cools an atom to the 3D motional ground-state starting from an initial temperature of 10-30~$\mu$K.}
The resulting pulse sequence, displayed in Fig.~\ref{pulse_sequence}, is composed of 4 groups of pulses which are repeated 5, 10, 10, and 15 times, respectively.
Groups 1, 2, and 3 have the form shown in Fig.~\ref{pulse_sequence}(a).
The radial pulses target the $n-1$ sideband with a duration that corresponds to a $\pi$-pulse for atoms in $n=3$.
In the axial direction, the pulse duration is chosen to maximise the Raman coupling for the desired sideband transition: $T = \pi/\eta_\mathrm{z}\Omega_\mathrm{z}(n_\mathrm{max},n_\mathrm{max}+\Delta n)$.
Group~1 applies pulses on the $n-4$ sideband with $n_\mathrm{max}=66$, group 2 applies pulses on the $n-3$ sideband with $n_\mathrm{max}=41$, and group 3 applies pulses on the $n-2$ sideband with $n_\mathrm{max}=21$.
Group 4 is the final set of pulses illustrated in Fig.~\ref{pulse_sequence}(b).
In the radial direction with the smaller trap frequency, x, we alternate the pulse duration between a $\pi$-pulse for $n=1$ or $n=3$ on the $n-1$ sideband.
In the axial direction, we alternate between pulses on the $n-2$ and $n-1$ sidebands with $n_\mathrm{max}=21$ and $n_\mathrm{max}=7$, respectively. 
The whole pulse sequence takes 45\,ms.
See \ref{appendix:pulse_sequence} for a table with full details of the pulse sequence.

\section{Ground-state Cooling in Separate Tweezers}
\label{sec:separate_tweezers}

\begin{figure}
    \centering
    \includegraphics[width=1\linewidth]{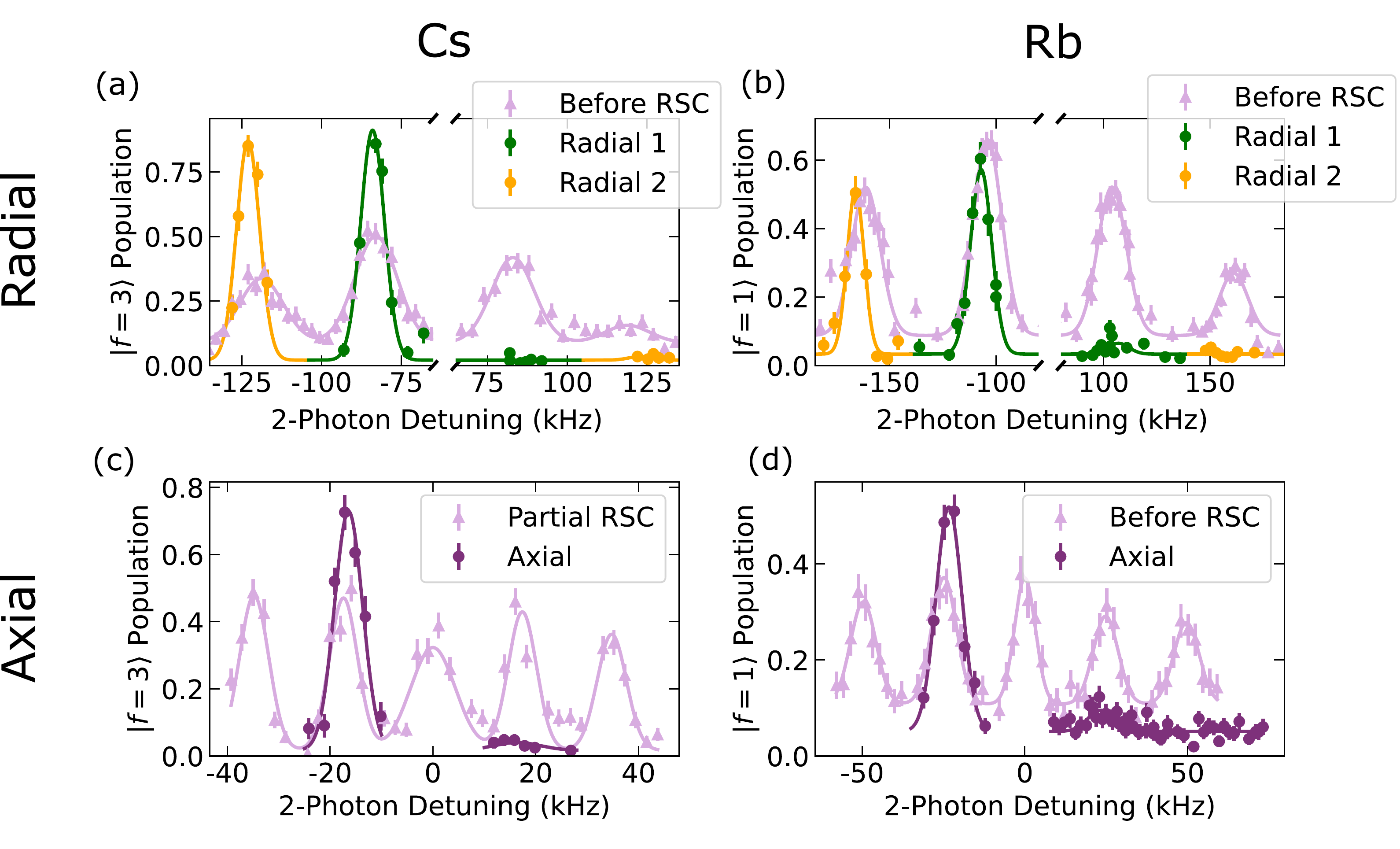}
    \caption{Raman sideband spectroscopy before and after the RSC pulse sequence. 
     In (a) and (b) a 100\,$\mu$s Tukey pulse with RB1+RB3 targets the radial directions. Trap asymmetry allows the two orthogonal radial axes to be resolved. Before RSC, plotted in lilac, both the BSB and RSB transitions are visible. After RSC the amplitudes of the RSB peaks vanish.
     In (c) and (d)  a 400\,$\mu$s Blackman-Harris pulse with RB1+RB4 targets the axial direction. Initially, the atom is outside of the LD regime, as is manifest in the visibility of higher order sideband transitions. The lilac line in (c) was taken after applying part of the RSC protocol on the $n-4$ and $n-3$ sidebands. The nonzero offsets in (a)--(d) are attributed to spin-changing photon-scattering events and imperfect state preparation and detection.
    }
    \label{sideband_spec}
\end{figure}

\noindent
The RSC pulse sequence is designed for high-fidelity preparation of single atoms in the motional ground state.
We confirm the effectiveness of the RSC pulse sequence using sideband thermometry in Fig.~\ref{sideband_spec}.
Assuming a thermal distribution, the ratio of the RSB and BSB peak amplitudes gives the {probability of occupying the motional ground state} $P(n=0) = 1 - A_{\mathrm{RSB}}/A_{\mathrm{BSB}}$ \cite{Diedrich1989}.
After RSC, we extract mean motional levels of {$\{ n_\mathrm{x}, n_\mathrm{y}, n_\mathrm{z}\}_\mathrm{Cs}  = \{0.000^{+0.014}_{-0.000}, 0.02^{+0.02}_{-0.02}, 0.03^{+0.03}_{-0.02}\}$ for Cs, and $\{ n_\mathrm{x}, n_\mathrm{y}, n_\mathrm{z}\}_\mathrm{Rb} = \{0.06^{+0.02}_{-0.02}, 0.00^{+0.04}_{-0.00}, 0.10^{+0.02}_{-0.02}\}$ for Rb.}
This corresponds to a 3D motional ground-state fraction of 0.95$^{+0.03}_{-0.04}$ for Cs in a 2~mK deep 938\,nm trap, and 0.86$^{+0.03}_{-0.04}$ for Rb in a 1.5~mK deep 814\,nm trap. The distribution after RSC is not thermal, but the sideband ratio method is still expected to give a sufficiently accurate estimate of the ground-state probability.
In the rest of this section, we examine the robustness of the cooling protocol, the dephasing of Rabi oscillations, and the application of the cooling protocol to atoms in an array of tweezers.

\subsection{Robustness of the Raman Sideband Cooling Protocol}
The RSC protocol is resilient against fluctuations in the {two-photon detuning and Rabi frequency}.
This {resilience} is achieved by applying enough pulses to saturate the ground-state probability.
The required number of pulses depends on the {detuning from two-photon resonance} and the pulse duration, but in the radial direction we can saturate the ground-state probability after $\sim 30$ pulses.
Our protocol uses 55 pulses on each radial axis so that the radial sideband detuning can drift by 6\,kHz with minimal effect on the final ground-state probability given the 20-30\,kHz Rabi frequencies.
Similarly, the axial sideband detuning must be set to within 2\,kHz.
This insensitivity to frequency offsets is promising for the extension of this scheme to larger arrays; normalisation of the array's trap intensities to within 10\% should result in comparable cooling performance across the array. 
The Raman beam powers fluctuate by $<3$\,\% ($<2$\,kHz change in light shift), such that the variations in the two-photon Rabi frequency and two-photon detuning are within the boundaries previously stated.
Furthermore, the reported 3D ground-state probabilities are achieved when simultaneously cooling Rb and Cs, demonstrating that crossover effects from the influence of the 780\,nm light on the Cs RSC, or 852\,nm light on the Rb RSC are negligible.
All together, our simultaneous RSC protocol is a robust method of achieving 3D ground-state probabilities of $>80\%$ for both species.

\begin{figure}
    \centering
    \includegraphics[width=1\linewidth]{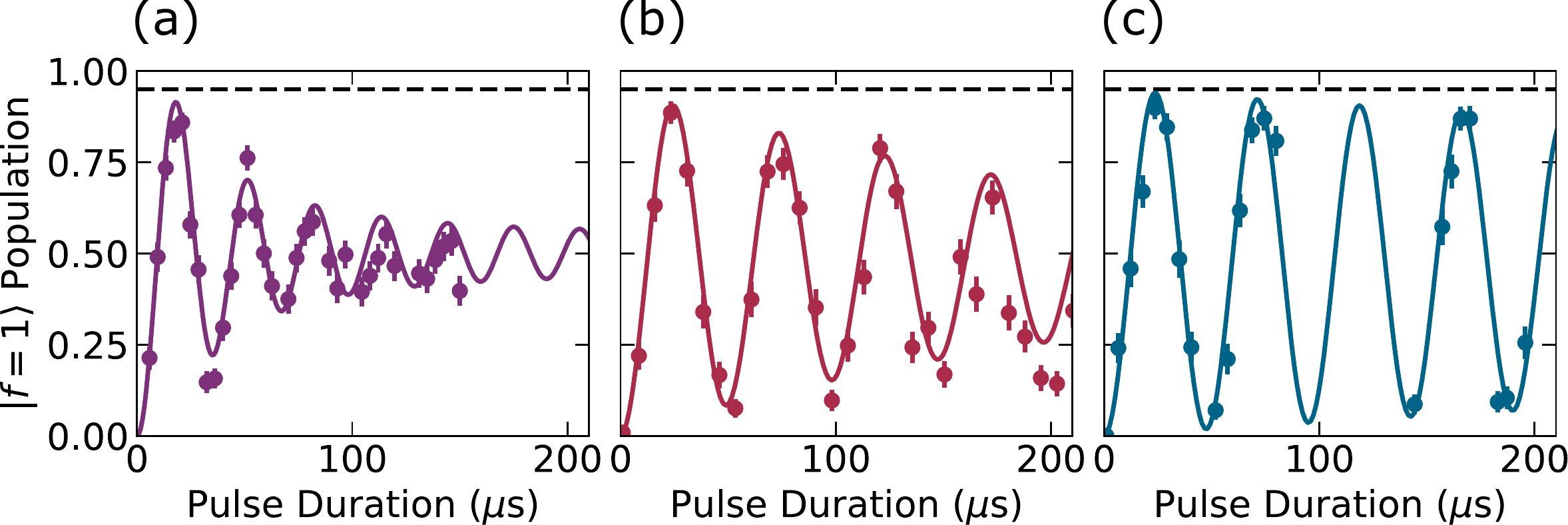}
    \caption{Carrier Rabi oscillations for a Rb atom with and without applying the RSC pulse sequence. (a) In an 814\,nm tweezer without applying RSC we fit a temperature of 25(2)\,$\mu$K and a Raman Rabi frequency of 32.8(3)\,kHz. Strong thermal dephasing is evident, so fitting a damped sine gives a 1/e time of 0.053(8)\,ms. (b) In an 814\,nm tweezer after applying RSC the thermal dephasing is removed and we fit a Rabi frequency of 20.3(1)\,kHz and a 1/e time of 0.25(4)\,ms. The remaining dephasing is due to the spread of differential light shifts from the tweezer. (c) The effect of differential light shifts is reduced by trapping in a 938\,nm tweezer, which is much further detuned from atomic transitions. We fit a damped sine with Rabi frequency 21.06(9)\,kHz and 1/e time 1.2(6)\,ms. The dashed line shows the expected state preparation and detection fidelity of 0.95.}
    \label{rabi_oscillations}
\end{figure}

\subsection{Dephasing of Rabi oscillations}
The effectiveness of the RSC protocol can be seen by examining Rabi oscillations on the Raman carrier transition before and after cooling. 
Fig.~\ref{rabi_oscillations} displays a measurement of carrier Rabi oscillations of a Rb atom using RB1+RB3 which can couple to the motion in the radial direction.
The carrier Rabi frequency can be evaluated using Eq.~\ref{eq:1} with $m=n$.
Without applying the RSC pulse sequence, the thermal distribution of motional levels leads to a distribution of Rabi frequencies and hence causes {dephasing}. 
This is evident in Fig.~\ref{rabi_oscillations}(a), taken in an 814\,nm tweezer without any RSC, where fitting a damped sine function allows us to extract a $1/e$ time of 0.053(8)\,ms.
{We can extract the temperature by instead fitting a sum over the Rabi oscillations from the different motional levels \cite{Kaufman2012};
\begin{equation}
    P_{F=1}(t) = \sum_n P_\mathrm{MB}(n, T) (1 - \cos (\Omega_\mathrm{r}(n,n) t))/2,
\end{equation}
where $P_\mathrm{MB}(n,T)$ is the Boltzmann probability of occupying motional state $n$ for a thermal distribution with temperature $T$.
We include the coupling to both radial axes using $\Omega_\mathrm{r}(n,n) = \Omega_\mathrm{x}(n,n) \Omega_\mathrm{y}(n,n) / \Omega_\mathrm{R}$ in order to extract the mean temperature.
We fit a Raman Rabi frequency of 32.8(3)\,kHz and a temperature of 25(2)\,$\mu$K. 
For comparison, the mean temperature from the radial sideband spectroscopy before RSC shown in Fig.~\ref{sideband_spec}(b) is 15(2)\,$\mu$K.
Typically, we expect a temperature of $\sim30\,\mu$K, as measured using the release and recapture method.} 

Fig.~\ref{rabi_oscillations}(b) shows the extended coherence time of carrier Rabi oscillations straight after applying the RSC protocol in the 814\,nm tweezer.
The fitted 1/e decay time is 0.25(4)\,ms.
In a precursor sideband thermometry measurement we measured a motional ground state probability of 0.86$^{+0.06}_{-0.06}$.
This high fidelity preparation into the motional ground state removes the effect of thermal dephasing. 
{We attribute the remaining dephasing {to a combination of factors stemming} from the bare diode laser source used for the tweezer. 
Firstly, the tweezer wavelength of 814\,nm is relatively near-detuned to the Rb D1 and D2 lines, leading to significant differential light shifts of $\sim15$\,kHz.
{Secondly,} we measure increased broadband intensity noise on the tweezer light after it has passed through the optical fibre which delivers light to the experiment and a subsequent polariser.
{We believe this} noise originates from multiple modes propagating in the fibre. 
{Together, these factors result in an enhanced spread of differential light shifts and increased dephasing.}} 
Note that this additional dephasing is also present in Fig.~\ref{rabi_oscillations}(a), meaning that the fitted temperature is likely an overestimate.
To confirm that the additional dephasing is a result of differential light shifts from the tweezer, Fig.~\ref{rabi_oscillations}(c) displays Rabi oscillations for Rb in a 938\,nm tweezer after applying the RSC protocol, where we extract a $1/e$ time of 1.2(6)\,ms.
{The 938\,nm tweezer has a similar level of intensity noise, and hence we measure fast dephasing of Rabi oscillations for a Cs atom.
Yet for Rb the dephasing due to the tweezer is greatly suppressed owing to the greater detuning from the Rb D1 and D2 lines.
We note that the additional dephasing associated with the bare diode laser can be removed by using a single {frequency} laser source. 
However, the robustness of our RSC protocol is demonstrated by the fact that we still prepare a single motional state with high fidelity despite the presence of this additional dephasing.}

\begin{figure}
    \centering
    \includegraphics[width=0.8\linewidth]{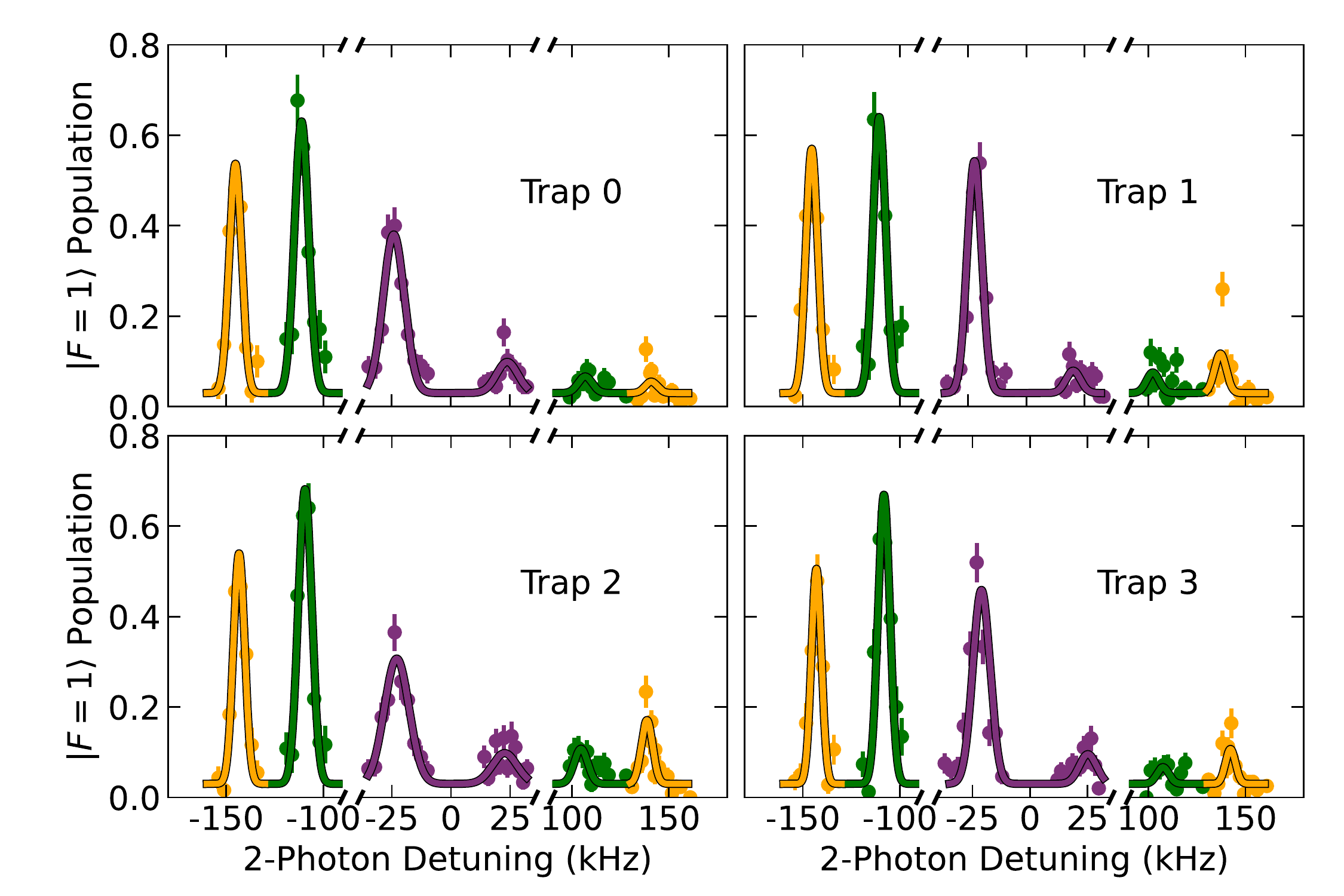}
    \caption{Sideband spectroscopy of an array of four optical tweezers that trap Rb atoms. Applying the RSC protocol simultaneously to the array achieves a probability of occupying the 3D motional ground state of $0.72^{0.05}_{0.05}$, $0.70^{0.07}_{0.10}$, $0.48^{0.08}_{0.12}$, and $0.67^{0.07}_{0.09}$ for trap 0, 1, 2, and 3 respectively. The tweezer wavelength is 817\,nm, different from the wavelength used in the rest of this paper. The traps are spaced 4\,$\mu$m apart. 
    }
    \label{rsc_array}
\end{figure}

\subsection{Cooling an Array}
In order to prepare an array of ultracold RbCs molecules, we must first prepare an array of atoms in the motional ground state. Fig.~\ref{rsc_array} displays sideband spectroscopy after applying the RSC protocol simultaneously to four tweezers with wavelength 817\,nm trapping Rb atoms in a 1D array with separation 4\,$\mu$m. The measured motional ground-state probability is  $0.72^{0.05}_{0.05}$, $0.70^{0.07}_{0.10}$, $0.48^{0.08}_{0.12}$, and $0.67^{0.07}_{0.09}$ for trap 0, 1, 2, and 3.  {Details of the generation of the array using a two-axis acousto-optic deflector (AOD) {added to our setup} can be found in \ref{appendix:array}}. The trap frequencies of the array were normalised to within $< 3\%$ of the mean value so that the two-photon detuning was within the bounds of efficient cooling by the RSC protocol; this level of array intensity normalisation has been achieved for arrays of over 100 atoms \cite{KeeslingThesis2021}. The motional ground-state preparation has a lower fidelity than a single trap because of two technical issues {that can be easily resolved in future work}. Firstly, the intensity noise on the tweezer light previously mentioned increases with the laser power. {Secondly, the polarisation purity measured straight after the two-axis AOD is reduced by a factor of between 10 - 70 across the array. This leads to a reduction in purity as measured using the atoms and to a decrease in the efficiency of the cooling.} Implementing a polariser after the AOD would restore the intended linear polarisation. Therefore, we conclude that our RSC protocol is suited for effective cooling of atoms in an array.

\section{Preparing Atom Pairs in the Relative Motional Ground State}
\label{sec:merging}
\begin{figure}
    \centering
    \includegraphics[width=1\linewidth]{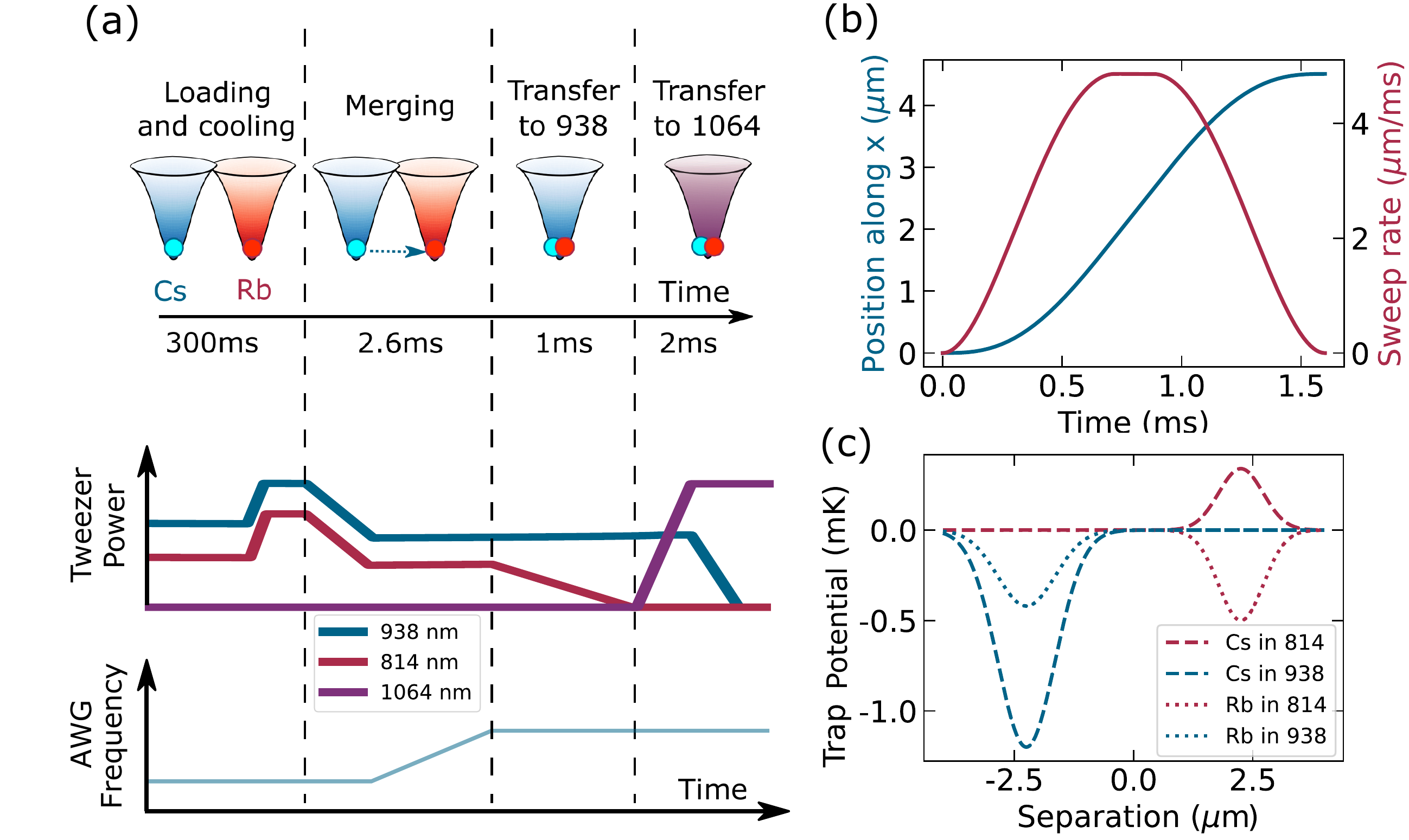}
    \caption{(a) The time sequence for merging atoms in the motional ground-state. 
    A Rb atom and a Cs atom are loaded, imaged, and cooled in separate tweezers. The 938\,nm tweezer is moved to the position of the 814\,nm tweezer and the 814\,nm tweezer is ramped off. Subsequently, both atoms are transferred into the 1064\,nm tweezer. In order to re-image the atoms, the process is reversed.
    To avoid pair loss caused by spin relaxation from most combinations of spin states, the sequence is followed while only loading one of the atomic species. 
    The lower time sequence depicts the relative powers of the tweezers. The arbitrary waveform generator (AWG) frequency shows when the 938\,nm tweezer is moved to overlap with the 814\,nm tweezer.
    (b) The hybrid minimum-jerk trajectory (blue) moving the 938\,nm tweezer a distance of 4.5\,$\mu$m has a sweep rate (red) that is constant for the central 10\,\% of the 1.6\,ms duration. 
    (c) The potentials experienced by a Rb atom (dotted line) or a Cs atom (dashed line) in the 814\,nm tweezer with power $P_{814}=0.64$\,mW (red) or the 938\,nm tweezer with power $P_{938}=3.8$\,mW (blue).}
    \label{merging_sequence}
\end{figure}

\noindent
Once a Rb and a Cs atom have been prepared in the motional ground-states of their respective tweezers, the next step on the route to creating molecules is to merge the traps in order to prepare an atom pair in a single optical tweezer. For molecule creation the atom pair must be in the relative motional ground state of the trap. 
Therefore, it is important that the transportation of the atoms and merging of the traps maintains the motional state. 
A balance must be found between merging slow enough to avoid motional excitation from the movement and not leaving excess time for photon scattering.
Our merging sequence, displayed in Fig~\ref{merging_sequence}(a), moves the Cs atom in the 938\,nm tweezer to the position of a Rb atom in a stationary 814\,nm tweezer, before transferring both atoms into a 1064\,nm tweezer at the same position.
We optimise the merging by decomposing the sequence to isolate the effects on each atom, as described below. We find that in order to avoid heating, we must carefully choose the trajectory and duration of movement for the 938\,nm tweezer in conjunction with the powers of both tweezers.

{First, we place a limit on the total duration of the merge by considering photon scattering. 
As previously mentioned, off-resonant scattering from the tweezers causes heating from photon recoil through Rayleigh scattering or {changes to the spin state through} Raman scattering which necessitates an {additional} OP step with associated photon recoil.}
We calculate a heating rate in the axial direction of 0.02\,quanta\,ms$^{-1}$ for Rb in the 814\,nm tweezer with a typical power of 1\,mW, and 0.004\,quanta\,ms$^{-1}$ for Cs in the 938\,nm tweezer for a typical power of 4\,mW.
These heating rates limit the merge duration to a few milliseconds. 

Secondly, we consider the limitations on the movement of the Cs atom in the 938\,nm tweezer alone. The position of the 938\,nm tweezer is dynamically controlled in the x-direction by an AOD, as previously described in Ref.~\cite{Brooks2021,Brooks2022}. 
Chirping the frequency of the RF signal driving the AOD translates the tweezer, but the duration of the sweep must be slow enough to avoid excitation.
To reliably keep the motional excitation $<0.01$ quanta in the direction of transport while adiabatically transporting a Cs atom 4.5\,$\mu$m in a 3.8\,mW 938\,nm tweezer, the duration must be $>0.11$\,ms \cite{Hucul2008}.
However, faster transport while maintaining the motional state is possible using shortcuts to adiabaticity \cite{Schulz2006,Couvert2008,Murphy2009, Chen2011, Torrontegui2011, Guery-Odelin2019, Lam2021}, assuming that the trap frequency is constant throughout the trajectory.

Unfortunately, the diffraction efficiency of the AOD has an oscillatory dependence on the driving frequency, which can cause resonant intensity modulation at certain sweep rates {(see \ref{appendix:sweep rate})}.
The resulting heating can be avoided by using a constant sweep rate where the intensity modulation is not resonant with the trap frequency. However, the linear chirp has significant jerk, which will heat the atom. Therefore, we form a hybrid minimum-jerk trajectory that starts and ends with a minimum-jerk function \cite{Liu2019}. 
Fig.~\ref{merging_sequence}(b) displays a trajectory with a duration of 1.6\,ms where the sweep rate is constant for the central 10\,\%. 
Using a trajectory {for 4.5\,$\mu$m of movement} with 10\,\% linear sweep, we can avoid the resonant intensity modulation when the sweep duration is $\sim1$\,ms or $\sim$1.6\,ms.

To maintain both atoms in the motional ground state, we must also consider the effect of combining the potential of the 938\,nm tweezer with that of the 814\,nm tweezer. We investigate this experimentally using the merging sequence displayed in Fig.~\ref{merging_sequence}(a).
We perform the experiment separately for each atomic species in order to avoid pair loss and interaction shifts complicating the interpretation of the sideband spectra used to measure any heating. The pair loss is a result of spin relaxation from all spin state combinations except when Rb is in $|f=1,m_f=1\rangle$ and Cs is in $|f=3,m_f=3\rangle$.
The final motional state of each atom is sensitive to both the trap depths (set by the tweezer powers, $P_{938}$ and $P_{814}$) and the overlap of the tweezers. We overlap the tweezers to within 100\,nm {in both radial directions} by pushing out a Cs atom from the 938\,nm or 1064\,nm tweezer using the repulsive potential of the 814\,nm tweezer. 
With the tweezers well overlapped, we observe that sweep durations longer than 1\,ms are required to avoid heating. Therefore we choose a sweep duration of 1.6\,ms which avoids resonant intensity modulation without leaving excess time for off-resonant scattering from the tweezer. All that remains is to choose the balance of the trap powers during the merge sequence. The Rb atom could be transferred to an excited motional state if the trap depths of the merging potentials are similar \cite{Kaufman2015}. Alternatively, when the combined potential experienced by Cs is very shallow, the close spacing of axial harmonic levels means there is a high probability of motional excitation. We explore the balance of trap powers by fixing $P_{938} = 3.8$~mW and varying $P_{814}$. 
The powers are adiabatically ramped in 1\,ms before merging the traps. The Rb atom starts to spill into the 938\,nm tweezer when the power ratio $P_{938}/P_{814} > 7$. In contrast, the trapping potential for the Cs atom vanishes when $P_{938}/P_{814} < 2$. In practice, we find that the heating of the Cs atom is more severe than that of the Rb atom, and a power ratio of $P_{938}/P_{814} \sim 6$ is required.

The last step of the merging sequence transfers the atoms to a 1064\,nm tweezer at the same position. {An atom in the 1064\,nm tweezer experiences a lower scattering rate} and therefore permits a greater precision for {the} sideband thermometry {used to assess the merging performance}. The wavelength is chosen in anticipation of the subsequent transfer to a molecular state, as for RbCs the molecular polarisability in the rovibrational ground state is similar to that of the {Feshbach molecule} at 1064\,nm \cite{Docenko2010, Vexiau2017, Blackmore2020}. 

\begin{figure}
    \centering
    \includegraphics[width=0.9\linewidth]{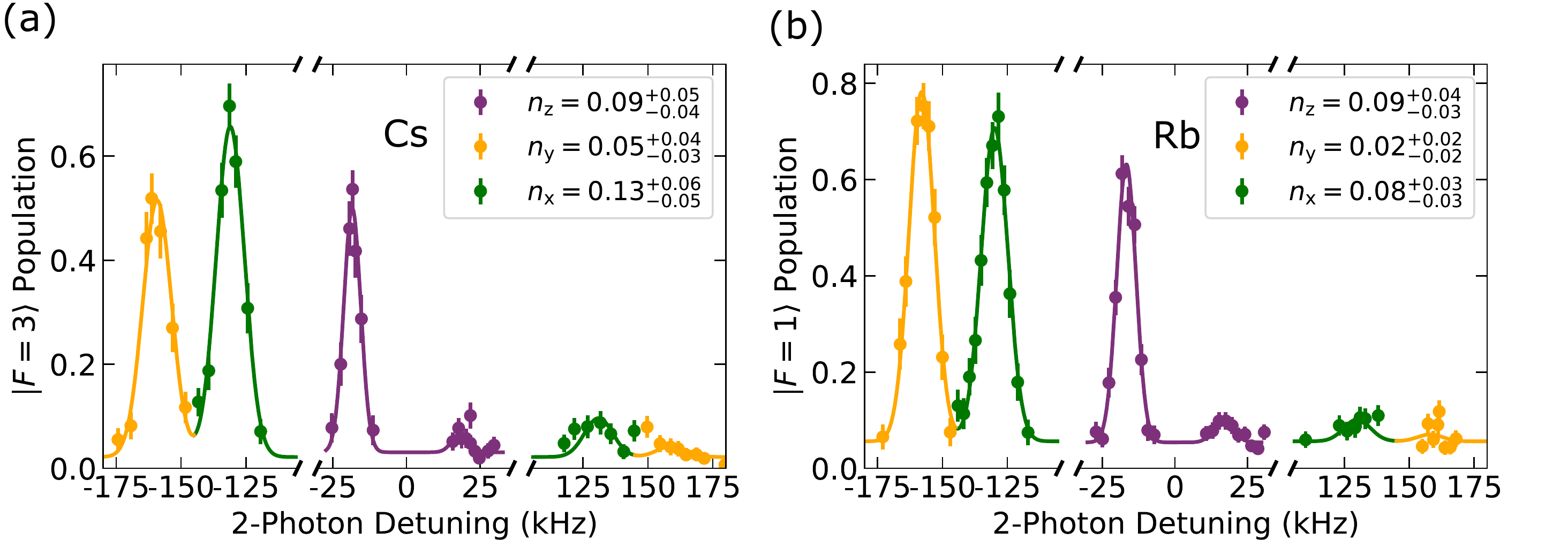}
    \caption{
    {Ground state occupation of a Rb-Cs atom pair in a common 1064\,nm tweezer. Raman sideband spectroscopy following the optimised merging sequence described in the text is used to measure a 3D ground-state probability of (a) 0.78$^{+0.05}_{-0.06}$ for Cs and (b) 0.83$^{+0.04}_{-0.04}$ for Rb.}
    }
    \label{merged_sideband_spec}
\end{figure}

Finally, we use the constraints previously discussed to determine parameters for merging a Rb atom in its motional ground-state with a Cs atom in its motional ground-state.
The tweezer powers are set at $P_{938} = 3.8$~mW and $P_{814} = 0.64$~mW. This power for the 938\,nm tweezer gives trap depths of $U^{938}_{\mathrm{Cs}} =1.2$~mK and $U^{938}_{\mathrm{Rb}}=0.42$~mK, whereas the 814\,nm tweezer provides a trap depth of $U^{814}_{\mathrm{Rb}}=0.50$~mK for Rb but functions as a barrier for Cs with $U^{814}_{\mathrm{Cs}}=-0.34~$mK. 
The separated potentials are plotted in Fig~\ref{merging_sequence}(c).
The trap powers are ramped in 1\,ms and the hybrid minimum-jerk trajectory has a duration of 1.6\,ms, with 10\,\% of the total duration being a linear sweep. 
The {whole} merging sequence is completed within 6\,ms.
Testing the sequence with a Cs atom, the mean motional levels extracted from the sideband spectroscopies in Fig.~\ref{merged_sideband_spec}(a) are $\{ n_\mathrm{x},~ n_\mathrm{y},~ n_\mathrm{z}\}_\mathrm{Cs} = \{0.13^{+0.04}_{-0.04},~ 0.05^{+0.03}_{-0.03},~ 0.09^{+0.04}_{-0.04}\}$. 
Repeating the measurements with a Rb atom, we extract mean motional levels of $\{ n_\mathrm{x},~ n_\mathrm{y},~ n_\mathrm{z}\}_\mathrm{Rb} = \{0.08^{+0.03}_{-0.03},~ 0.02^{+0.02}_{-0.02},~ 0.09^{+0.03}_{-0.03}\}$ from the sideband spectroscopies in Fig.~\ref{merged_sideband_spec}(b). This corresponds to a 3D ground-state probability of 0.78$^{+0.05}_{-0.06}$ for Cs and 0.83$^{+0.04}_{-0.04}$ for Rb. 
While the decrease in ground-state probability of the Rb atom is consistent with heating from tweezer scattering, the optimised parameters have not completely prevented motional excitation of the Cs atom. We expect an increase of 0.028 quanta due to photon scattering from the tweezer. Yet, the mean motional level in the x-direction has increased by 0.13$^{+0.04}_{-0.04}$ quanta. 
Most likely this is a consequence of heating from the resonant modulation of the AOD's diffraction efficiency, and the heating is greater than measured with just Cs in the 938\,nm tweezer because the trap frequencies are modified when the two tweezer potentials combine.

The proportion of atoms in the relative motional ground state can be estimated using \cite{Zhang2020} 
\begin{equation}
    P(n_{\mathrm{rel}}=0) = \frac{P(n_{\mathrm{Rb}}=0) P(n_{\mathrm{Cs}}=0)}{1 - \frac{m_{\mathrm{Cs}}}{m_{\mathrm{Cs}}+m_{\mathrm{Rb}}}\frac{\bar{n}_{\mathrm{Cs}}}{\bar{n}_{\mathrm{Cs}}+1} - \frac{m_{\mathrm{Rb}}}{m_{\mathrm{Cs}}+m_{\mathrm{Rb}}}\frac{\bar{n}_{\mathrm{Rb}}}{\bar{n}_{\mathrm{Rb}}+1}}.
\end{equation}
The derivation of this formula makes three assumptions. 
Firstly, that the centre-of-mass and relative motion are separable. 
In the 1064\,nm tweezer the Rb and Cs trap frequencies are similar, satisfying $\omega_\mathrm{Cs} \approx 1.08 \omega_\mathrm{Rb}$, such that the centre-of-mass and relative motion are approximately separable.
Secondly, that the distribution of motional levels follows a Boltzmann distribution. 
Thirdly, the derivation ignores interactions between the atoms.
While the interactions between the atoms will shift the energy eigenstates \cite{Hood2020}, the motional ground state that we prepare is adiabatically connected to $|n_\mathrm{rel}=0\rangle$ as the interaction strength is tuned to zero, provided that we can ignore the unlikely effects of trap-induced shape resonances \cite{Krych2009, Fogarty2011}.
Therefore, the probability of preparing an atom pair in the relative motional ground state, which becomes $ |f_{\mathrm{Rb}}=2,m_{f,\mathrm{Rb}}=2;f_{\mathrm{Cs}}=4, m_{f,\mathrm{Cs}}=4;n_{\mathrm{rel}}=0\rangle$ in the limit of weak interactions, is $P(n_{\mathrm{rel}}=0) = 0.81^{+0.08}_{-0.08}$ following our optimised sequence.
Using adiabatic rapid passage with a microwave field \cite{Bloch1946, Camparo1984} or a Raman carrier $\pi$-pulse before merging, we can transfer to the desired hyperfine spin state with $>90$\% fidelity. Then $>60$\% of the initial atom pairs will be transferred into the $|1,1;3,3;n_{\mathrm{rel}}=0\rangle$ state. This is the state from which magnetoassociation to a molecular state can be achieved following a similar scheme employed with bulk mixtures of Rb and Cs \cite{Takekoshi2012, Koppinger2014}.

\section{Conclusion}
\label{sec:conclusion}
\noindent
We have demonstrated a pulsed Raman sideband cooling sequence that simultaneously cools a Rb atom and a Cs atom to the 3D motional ground-states of their respective traps with high fidelity. Since the atoms start outside of the Lamb-Dicke regime, the sequence starts with axial sideband transitions on the fourth red sideband, $\Delta n=-4$, to ensure the portion of the population distribution initially in high motional levels is also cooled.
Using sideband thermometry we measure a 3D ground-state fraction of 0.86$^{+0.03}_{-0.04}$ for Rb in a 814\,nm tweezer, and 0.95$^{+0.03}_{-0.04}$ for Cs in a 938\,nm tweezer. 
The scalability and robustness of the cooling protocol is promising for expansion to arrays of tweezer traps, and we demonstrate simultaneous cooling of Rb atoms in an array of four tweezers. {Application of our RSC protocol to dual-species arrays of Rb and Cs atoms \cite{Singh2022} would allow initialisation of both species in the motional ground-state, improving the fidelity of tweezer-based quantum logic gates \cite{Bluvstein2022,Graham2022}. In addition, this would realise an ideal system for implementing the mid-circuit measurements \cite{Beterov2015} that are necessary for quantum error correction schemes \cite{Auger2017}.}

{We have also demonstrated that the Rb and Cs atoms can be transferred into the same 1064\,nm tweezer trap with little heating;} we measure 3D ground-state fractions of 0.83$^{+0.04}_{-0.04}$ for Rb and 0.78$^{+0.05}_{-0.06}$ for Cs {in the combined trap}. The far-detuned 1064\,nm tweezer has the benefit of a reduced scattering rate. {This trap is also well suited to molecule formation as} at this wavelength the polarisability of the Feshbach and ground molecule states are similar \cite{Blackmore2020}. The inferred occupation fraction of the relative motional ground-state $|f_{\mathrm{Rb}}=2,m_{f,\mathrm{Rb}}=2;f_{\mathrm{Cs}}=4, m_{f,\mathrm{Cs}}=4;n_{\mathrm{rel}}=0\rangle$ is 0.81$^{+0.08}_{-0.08}$. With efficient microwave or Raman carrier transitions we can populate the $|1,1;3,3;n_{\mathrm{rel}}=0\rangle$ state with efficiency $>60\%$. 
The natural next step is to establish magnetic field control that allows sweeping of the magnetic field across an interspecies Feshbach resonance to form weakly bound RbCs molecules {\cite{Takekoshi2012, Koppinger2014}}, paving the way to the production of single polar molecules in the rovibrational ground state using STIRAP {\cite{Molony2014, Takekoshi2014}}. 

\section{Acknowledgements}
The authors thank L A McArd and A Hunter for designing and building the DDS AOM driver circuit.
The authors thank R Sawant for initial ideas in simulating RSC, S White for simulations of merging traps, and A Alampounti for developing software control of the AWG.
This work made use of the facilities of the Hamilton HPC Service of Durham University.
This work was supported by U.K. Engineering and Physical Sciences Research Council (EPSRC) Grant EP/P01058X/1 and Durham University. 

\section{Data Availability Statement}
The data that support the findings of this study are openly available at the following URL/DOI: 10.15128/r1f4752g787.
\vspace{20pt}

\bibliographystyle{iopart-num.bst}
\bibliography{references}

\appendix

\section{Coupling Hyperfine Sub-levels with a Raman Transition}
\label{appendix:dme}
\begin{figure}
    \centering
    \includegraphics[width=1\linewidth]{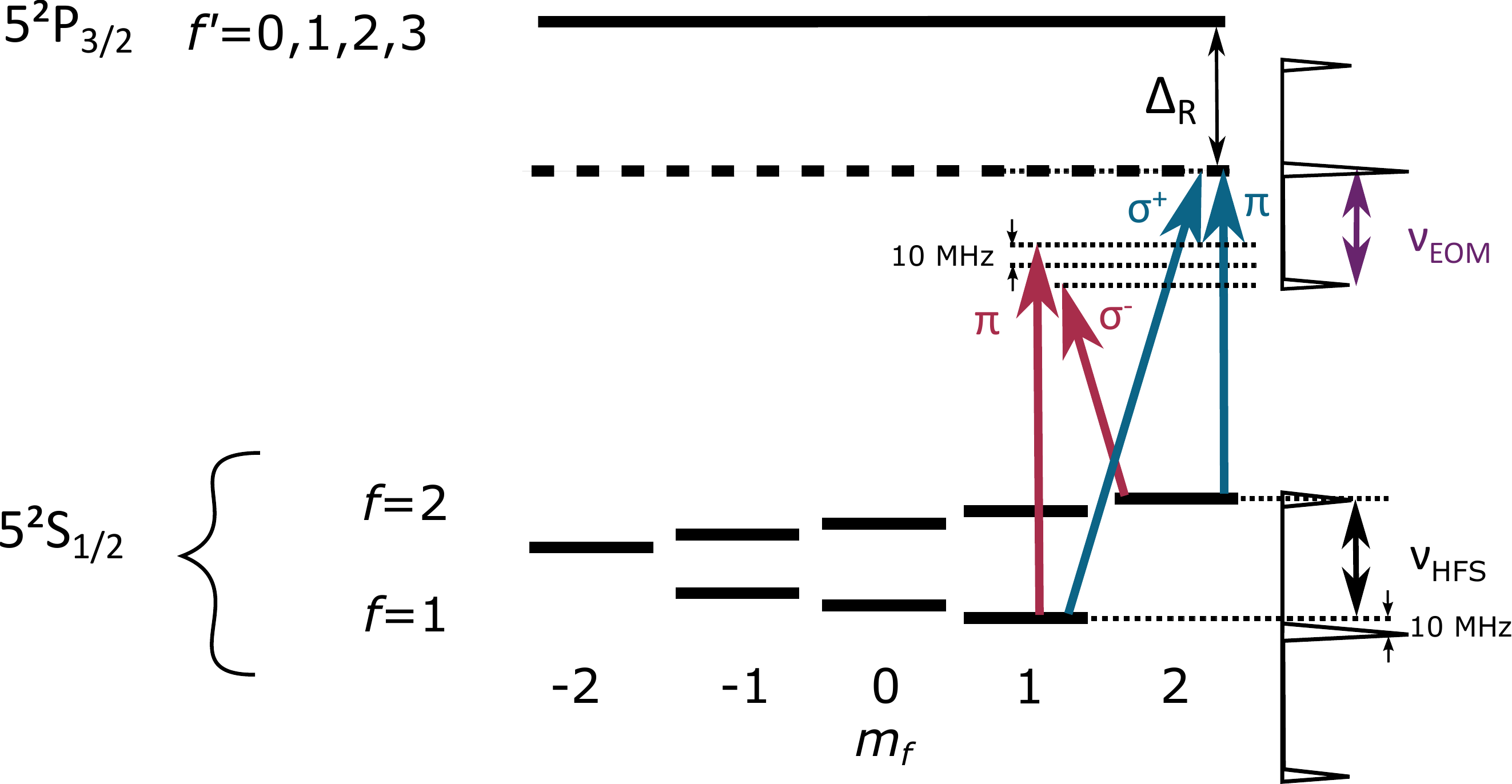}
    \caption{Simplified diagram of the hyperfine structure of a Rb atom showing the desired two-photon Raman transition and an undesired Raman transition. The hyperfine structure for Cs has different values for the total angular momentum quantum number $f$. Blue: a two-photon Raman transition between two of the spin-stretched states is possible when one Raman beam drives a $\sigma+$ transition and the other drives a $\pi$ transition. The energy difference is added to one of the Raman beams using the upper sideband of an EOM. Red: if the circularly polarised Raman beam has residual light of the opposite handedness, it can drive a two-photon Raman transition with the lower EOM sideband that destructively interferes with the desired transition. We suppress this possibility by offsetting the EOM frequency by 10\,MHz and using the AOM frequencies to bring the desired transition back onto two-photon resonance.}
    \label{hyperfine_structure}
\end{figure}

Fig.~\ref{hyperfine_structure} displays two two-photon Raman transitions between the spin-stretched hyperfine states used in this paper which are allowed by dipole selection rules.
The desired transition, in blue, occurs via a single virtual excited state.
The other transition, in red, is possible through several virtual excited states and destructively interferes with the desired transition. 
Therefore, we suppress the undesired transition by using an 1000~:~1 polariser and a quarter waveplate with a retardance of 0.267 waves at 780\,nm, and 0.243 waves at 852\,nm.
Assuming linearly polarised incident light and that the waveplate angle is set to within 1 degree, this maintains a polarisation purity of 500~:~1 (calculated using Jones matrices \cite{Jones1941}). 
Secondly, we offset the EOM frequency by 10\,MHz, $\nu_\mathrm{EOM} = \nu_\mathrm{HFS} + 10$\,MHz, such that the undesired transition is $-20$\,MHz detuned from two-photon resonance.
AOM1 controls the frequency of the laser beam driving $\sigma^\pm$ transitions, and AOM2 controls the frequency of the laser beam driving $\pi$ transitions.
Achieving two-photon resonance with the two laser beams marked by blue lines in Fig.~\ref{hyperfine_structure} implies that $\nu_\mathrm{AOM1} = \nu_\mathrm{AOM2} + \nu_\mathrm{HFS} - \nu_\mathrm{EOM} = \nu_\mathrm{AOM2} - 10$\,MHz. 
Note that the upper EOM sideband is used for the blue line driving $\sigma^+$ transitions, whereas the lower EOM sideband is used for the red line driving $\sigma^-$ transitions.
In both cases the carrier frequency and other EOM sideband are detuned from two-photon resonance by at least the ground-state hyperfine splitting.
The detuning of the laser beams marked by red lines in Fig.~\ref{hyperfine_structure} is then given by $\nu_\mathrm{AOM1} - \nu_\mathrm{EOM} + \nu_\mathrm{HFS} - \nu_\mathrm{AOM2} = -20$\,MHz.

\section{Simulations of Raman Sideband Cooling}
\label{appendix:simulation}
\noindent
The design of the pulse sequence discussed in Section~\ref{pulse_sequence} was informed by simulations which solve the Lindblad master equation \cite{Cirac1994} numerically using QuTiP \cite{Johansson2012}.
An insightful model is made using the simplifying assumptions that the atom is in a harmonic potential interacting either with the coherent drive of the Raman beams, or a dissipative term from the OP beams.
The time-dependent Hamiltonian includes the coherent terms from the harmonic confinement, the internal atomic spin state, and an interaction term from coupling to the Raman beam:
\begin{equation}
    H = \sum_n{n \hbar\omega|n\rangle\langle n|}\otimes\mathcal{I} - \mathcal{I}\otimes \frac{\hbar\delta\sigma_z}{2} + H_\mathrm{R}.
    \label{eq:Jaynes-Cummings}
\end{equation}
Here $\omega$ is the trap frequency that separates the motional levels $|n\rangle$, $\delta$ is the 2-photon detuning, $\sigma_\mathrm{z}$ is the third Pauli spin operator, and $H_\mathrm{R}$ is the Raman interaction Hamiltonian.
We drop the ground state energy $\frac{\hbar\omega}{2}|\downarrow; n=0 \rangle\langle\downarrow; n=0|$ which provides an arbitrary offset.
The interaction Hamiltonian from the Raman beams couples the motional levels with the internal spin states \cite{Meekhof1996}:
\begin{equation}
    H_\mathrm{R} = \frac{\hbar\Omega_\mathrm{R}}{2} \left(e^{i\eta (a + a^\dagger)} \otimes \sigma^+ + e^{-i\eta (a + a^\dagger)} \otimes \sigma^- \right).
    \label{eq:RamanHamiltonian}
\end{equation}
Where the momentum kick is $\hbar\Delta k = \hbar|\mathbf{k_\pi - k_\sigma}|$, with $\Delta k \hat{x} = \eta(a + a^\dagger)$ for Raman LD parameter $\eta$.
$\Omega_\mathrm{R}$ is the Raman Rabi frequency.
We introduce the spin-raising operator $\sigma^+ = \sigma_x + i\sigma_y$.
For the dissipative OP term, we assume that each OP step scatters 3 photons \cite{Liu2019}, giving
\begin{equation}
    \gamma_\mathrm{OP} = \sqrt{\Gamma_\mathrm{OP}}\left( e^{i\eta_\mathrm{OP}(a + a^\dagger)} \right)^3 \otimes \sigma^- .
    \label{eq:OP-Hamiltonian}
\end{equation}
$\sigma^- = \sigma_x - i\sigma_y$ is the spin-lowering operator.
$\Gamma_\mathrm{OP}$ is the scattering rate of the OP beams, and the recoil from the OP photons defines the OP LD parameter $\eta_\mathrm{OP}$.
We have simplified the discussion to a single dimension.
The generalisation to 3D is trivial, but since the directions are independent we retain the simplicity of using 1D and analyse each direction separately.
{An exception is the radial Rabi oscillations in Fig.~\ref{rabi_oscillations}, where we consider the coupling to both radial axes, assuming that the position operators are separable: 
\begin{equation}
    \begin{array}{ll} 
    \frac{\Omega(n_\mathrm{x},m_\mathrm{x}; n_\mathrm{y},m_\mathrm{y})}{\Omega_\mathrm{R}} &= |\langle n_\mathrm{x}, n_\mathrm{y} | \exp{(i\mathbf{\Delta k \cdot \hat{r}})} |m_\mathrm{x}, m_\mathrm{y} \rangle| \\
    &=|\langle n_\mathrm{x}|\exp{(j\eta_\mathrm{x}(a+a^{\dagger}))} |m_\mathrm{x} \rangle \langle n_\mathrm{y}|\exp{(i\eta_\mathrm{y}(a+a^{\dagger}))} |m_\mathrm{y} \rangle|   \\
    &= \Omega_\mathrm{x}(n_\mathrm{x},m_\mathrm{x}) \Omega_\mathrm{y}(n_\mathrm{y},m_\mathrm{y})/ \Omega_\mathrm{R}.
    \end{array}
\end{equation}
The last line follows from Eq.~\ref{eq:1}, and the fitted line in Fig.~\ref{rabi_oscillations} assumes $n_\mathrm{x} = n_\mathrm{y}$ in order to extract the mean temperature.}

The simulation of a full pulse sequence runs in several minutes provided that the basis of harmonic levels included is sufficiently small.
For our typical starting temperatures, the thermal distribution contains 99\,\% of the population within motional levels $n<21$ for the radial direction with $\langle n \rangle = 4$ and $\eta\sim 0.15$.
However, in the axial direction with $\langle n \rangle = 10$ and $\eta\sim 0.3$ we must include up to $n=65$ in order to represent 99\,\% of the population.

\section{3D Raman Sideband Cooling Pulse Sequence}
\label{appendix:pulse_sequence}
Table~\ref{table:RSC_pulses} gives full details of the transition frequencies and Rabi frequencies used in the different stages of the RSC pulse sequence.

\begin{table}[tb]
      \centering
      \begin{tabular}{ |c|c|c|c|c|c| } 
     \hline
     Group 1 &  axial $\Delta n=-4$ & 5 repetitions & & \\ 
     \hline
     \thead{Cs Sideband Transition\\ Frequency (kHz)} & \thead{Rb Sideband Transition\\ Frequency (kHz)} & \thead{Axis} & \thead{Rb Raman Rabi \\Frequency (kHz)} & \thead{Duration ($\mu$s)} \\
     \hline
     68 & 100 & Axial & 4.5 & 272 \\
     76 & 110 & Radial 1 & 21 & 70 \\
     68 & 100 & Axial & 4.5 & 272 \\
     112 & 170 & Radial 2 & 27 & 60 \\
     \hline
     \hline
     Group 2 & axial $\Delta n=-3$ & 10 repetitions & & \\ 
     \hline
     \thead{Cs Sideband Transition\\ Frequency (kHz)} & \thead{Rb Sideband Transition\\ Frequency (kHz)} & \thead{Axis} & \thead{Rb Raman Rabi \\Frequency (kHz)} & \thead{Duration ($\mu$s)} \\
     \hline
     51 & 75 & Axial & 4.5 & 251 \\
     76 & 110 & Radial 1 & 21 & 70 \\
     51 & 75 & Axial & 4.5 & 251 \\
     112 & 170 & Radial 2 & 27 & 60 \\
     \hline
     \hline
     Group 3 & axial $\Delta n=-2$ &  10 repetitions & & \\ 
     \hline
     \thead{Cs Sideband Transition\\ Frequency (kHz)} & \thead{Rb Sideband Transition\\ Frequency (kHz)} & \thead{Axis} & \thead{Rb Raman Rabi \\Frequency (kHz)} & \thead{Duration ($\mu$s)} \\
     \hline
     34 & 50 & Axial & 4.5 & 224 \\
     76 & 110 & Radial 1 & 21 & 70 \\
     34 & 50 & Axial & 4.5 & 224 \\
     112 & 170 & Radial 2 & 27 & 60 \\
     \hline
     \hline
     Group 4 & axial $\Delta n=-2, -1$ &  15 repetitions & & \\ 
     \hline
     \thead{Cs Sideband Transition\\ Frequency (kHz)} & \thead{Rb Sideband Transition\\ Frequency (kHz)} & \thead{Axis} & \thead{Rb Raman Rabi \\Frequency  (kHz)} & \thead{Duration ($\mu$s)} \\
     \hline
     34 & 50 & Axial & 4.5 & 224 \\
     76 & 110 & Radial 1 & 21 & 100 \\
     17 & 25 & Axial & 4.5 & 188 \\
     112 & 170 & Radial 2 & 27 & 60 \\
     17 & 25 & Axial & 4.5 & 281 \\
     76 & 110 & Radial 1 & 21 & 100 \\
     34 & 50 & Axial & 4.5 & 224 \\
     76 & 110 & Radial 1 & 21 & 70 \\
     17 & 25 & Axial & 4.5 & 188 \\
     112 & 170 & Radial 2 & 27 & 60 \\
     \hline
     \end{tabular}
      \caption{ The pulse sequence used for simultaneous RSC of Rb and Cs. The sideband transition frequency is the difference between the frequency of the carrier transition and the desired sideband transition. The quoted Raman Rabi frequencies are mean values. For a convolution of a Tukey profile and a square pulse, as used to pulse shape the radial directions, this is different from the peak Rabi frequency by a factor of 1.22. For a Blackman-Harris profile the mean Rabi frequency differs from the peak Rabi frequency by a factor of 2.79. The duration of the OP pulses is 15\,$\mu$s and the entire pulse sequence lasts 45.05\,ms
      }
      \label{table:RSC_pulses}
\end{table}

\section{Generation of an Array of Four Tweezers}
\label{appendix:array}

We generate an array of tweezer traps by deflecting the tweezer laser in several directions using a 2D AOD (AA Opto Electronic DTSXY-400-810).
The traps originate from the same laser at a wavelength of 817\,nm, different to the wavelength used for Rb throughout the rest of this paper. 
The AOD is driven by an arbitrary waveform generator (Spectrum Instrumentation M4i.6631-x8) set to a sample rate of 1024\,MS/s.
One channel is used for horizontal deflection, and a second channel is used for vertical deflection.
We drive the horizontal AOD with a signal composed of four RF tones equally spaced across a range of 6\,MHz, which creates four tweezer traps in the atom plane with a 4\,$\mu$m separation between each trap.
The frequencies are chosen so that there are an integer number of periods within the data loaded onto the card, in order to avoid noise associated with phase slips when looping segments of data loaded onto the AWG.
The phase of each tone is optimised to reduce interference from the mixing of sum and difference tones generated by the nonlinear response of the amplifier and AOD \cite{Schroeder1970,Levine2021thesis}.

We homogenise the intensities of the traps across the array in order to ensure the Raman transitions during RSC are on two-photon resonance.
The aforementioned phase optimisation is an essential first step, after which driving each tone with the same RF power results in optical powers differing by $\sim20$\,\%.
Then we change the amplitudes of the RF tones until the powers measured on a CCD camera placed soon after the AOD are normalised to within $3$\,\%.
We then confirm the axial trap frequencies are normalised to within $3$\,\% using a parametric heating measurement.
Balancing the trap intensities means that the tweezer light shifts are the same across the array.

\section{Etaloning in an Acousto-Optic Deflector Crystal}
\label{appendix:sweep rate}

There is a technical issue that can occur when using an AOD to move an optical tweezer, which can cause severe heating of the atom during transport. 
The problem originates from the AOD crystal acting as an etalon \cite{Liu2019thesis}.
An input RF signal drives a piezo-electric transducer to create an acoustic travelling wave in the AOD crystal, which periodically modifies the refractive index to deflect an input laser beam through Bragg diffraction.
However, reflections of the acoustic wave from the opposite edge of the crystal will interfere and create a standing wave if the wavelength matches the cavity length.
The consequence is that the diffraction efficiency of the AOD oscillates with a period given by the free spectral range of the cavity, $c/2L$, where $c$ is the speed of sound in the cavity and $L$ is the cavity length.
For our AOD (ATD-1803DA2.850 from IntraAction) with speed of sound 4200~m\,s$^{-1}$, we measure a cavity free spectral range of 175\,kHz.
This implies a cavity length of 12\,mm, which roughly matches the crystal dimensions.
Typically the reflections are small and so the oscillations in diffraction efficiency are only a few per cent \cite{Liu2019}. 
The interference from reflection can be mitigated by angling the far edge of the crystal.
However, our AOD is not angle-cut, and the relative diffraction efficiency modulation is between 0.1-0.5\% ({the variation in the modulation is dependent on the RF drive}).
When the driving frequency is swept to move the deflected beam, the intensity of the beam follows the oscillations in diffraction efficiency.
{If this intensity modulation is at double the trap frequency, there would be parametric heating of the atom \cite{Savard1997}.
However, we find experimentally that the most significant heating occurred when the intensity modulation is at the trap frequency, rather than double the trap frequency.
This suggests that the heating is from pointing noise rather than intensity noise \cite{Savard1997}, and that there is a coupling between the tweezer intensity and the position of the atom.
This coupling could arise from the presence of strong vector light shifts as discussed in the main text.
}
To minimise the heating from this effect, we avoid the sweep rates given by:
\begin{equation}
    v_i = \frac{c}{2L}\omega_{\mathrm{trap}, i}\frac{dx}{df},
\end{equation}
where $\omega_{\mathrm{trap}, i}$ is the trapping frequency along the axis $i$, and $\frac{dx}{df}=0.322\,\mu$m\,MHz$^{-1}$ is the proportionality constant between the frequency of the RF input and the movement of the optical tweezer in the atom plane.
Using the radial trap frequencies of 64\,kHz and 92\,kHz for a 938\,nm tweezer power of 3.8~mW we calculate that sweep rates of $3.6\,\mu$m\,ms$^{-1}$ or $5.2\,\mu$m\,ms$^{-1}$ cause resonant intensity modulation.

The trajectory used to move the optical tweezer is chosen to avoid resonant intensity modulation and heating from jerk.
Varying the driving frequency of the AOD with a constant sweep rate makes it easy to avoid resonant intensity modulation. 
However, the sudden change in acceleration at the start and end of the sweep can cause motional excitation of the atom being transported.
Using a minimum-jerk trajectory \cite{Shadmehr2005} resolves the heating associated with jerk.
However, the sweep rate varies throughout the trajectory and is likely to scan over a range that causes parametric heating.
A compromise can be found using a hybrid minimum-jerk trajectory, which we define as in \cite{Liu2019}: 
\begin{equation}
    x_\mathrm{min-jerk}(t, d, T) = d \left[ 10 \left(t/T \right)^3 - 15 \left(t/T\right)^4 + 6 \left(t/T\right)^5 \right].
\end{equation}

\begin{equation}
    x_\mathrm{hybrid} = \left\{
    \begin{array}{ll} 
    x_\mathrm{min-jerk}(t,2\Delta f, 2\Delta t) & \mathrm{for~} 0 \leq t \leq \Delta t \\
    \frac{15}{4} \frac{\Delta f}{2\Delta t} & \mathrm{for~} \Delta t < t < T - \Delta t \\
    x_\mathrm{min-jerk}(t-T+2\Delta t,2\Delta f, 2\Delta t) + \alpha T \frac{15}{4} \frac{\Delta f}{2\Delta t} & \mathrm{for~} T-\Delta t < t \leq T
    \end{array}
    \right\}
\end{equation}
\begin{equation}
    \Delta f = \frac{d}{\left(2 + \frac{15\alpha}{4(1-\alpha)}\right)}
\end{equation}
Here, $\Delta f$ is the distance travelled during the minimum-jerk portion of the trajectory, and $\Delta t = T(1-\alpha)/2$ is the time elapsed during the minimum-jerk portion. 
The fraction of the total duration which follows the linear trajectory ($\alpha=0$ for fully minimum-jerk and $\alpha=1$ for fully linear) is colloquially named the hybridicity.
Both the duration of the sweep, $T$, and the hybridicity, $\alpha$, determine whether we scan across a bad sweep rate.
{In this paper trajectories with a hybridicity of 0.1 were used to move a distance of 4.5\,$\mu$m (a frequency chirp of 14\,MHz). This implies that the intensity modulation will be resonant with the trap frequencies $\{\omega_{\mathrm{x}}, \omega_{\mathrm{y}} \}^{938}_\mathrm{Cs} = \{64, 92\}$\,kHz for sweep durations of \{2.1, 1.5\}\,ms respectively.
We have demonstrated that the motional ground state can be maintained with high fidelity when merging a Cs atom with a Rb atom using a trajectory with hybridicity 0.1 and duration 1.6\,ms. 
Future work might explore using a different hybridicity and sweep duration.}

\end{document}